\documentclass[sigconf]{acmart}
\AtBeginDocument{%
  \providecommand\BibTeX{{%
    \normalfont B\kern-0.5em{\scshape i\kern-0.25em b}\kern-0.8em\TeX}}}

\usepackage[ruled,linesnumbered]{algorithm2e}
\usepackage{bm}

\copyrightyear{2022}
\acmYear{2022}
\setcopyright{rightsretained}
\acmConference[MobiHoc '22]{The Twenty-second International
Symposium on Theory, Algorithmic Foundations, and Protocol Design for
Mobile Networks and Mobile Computing}{October 17--20, 2022}{Seoul,
Republic of Korea}
\acmBooktitle{The Twenty-second International Symposium on Theory,
Algorithmic Foundations, and Protocol Design for Mobile Networks and
Mobile Computing (MobiHoc '22), October 17--20, 2022, Seoul, Republic
of Korea}\acmDOI{10.1145/3492866.3549719}
\acmISBN{978-1-4503-9165-8/22/10}

\newtheorem{theorem}{{\bf Theorem}}

\newtheorem{lemma}{{\bf Lemma}}

\begin{document}

\title{
Optimizing Age of Information with Correlated Sources
}

\author{Vishrant Tripathi}
\email{vishrant@mit.edu}
\affiliation{\institution{Massachusetts Institute of Technology} 
\country{USA}
}
\author{Eytan Modiano}
\email{modiano@mit.edu}
\affiliation{\institution{Massachusetts Institute of Technology} 
\country{USA}
}

\renewcommand{\shortauthors}{Tripathi and Modiano}

\begin{abstract}
We develop a simple model for the timely monitoring of correlated sources over a wireless network. Using this model, we study how to optimize weighted-sum average Age of Information (AoI) in the presence of correlation. First, we discuss how to find optimal stationary randomized policies and show that they are at-most a factor of two away from optimal policies in general. Then, we develop a Lyapunov drift-based max-weight policy that performs better than randomized policies in practice and show that it is also at-most a factor of two away from optimal. Next, we derive scaling results that show how AoI improves in large networks in the presence of correlation. We also show that for stationary randomized policies, the expression for average AoI is robust to the way in which the correlation structure is modeled. Finally, for the setting where correlation parameters are unknown and time-varying, we develop a heuristic policy that adapts its scheduling decisions by learning the correlation parameters in an online manner. We also provide numerical simulations to support our theoretical results. 
\end{abstract}




\maketitle

\section{Introduction}
\label{sec:intro}

Monitoring, estimation, and control of systems are fundamental  and well studied problems. Many emerging applications involve performing these tasks over communication networks. Examples include: sensing for IoT applications, control of robot swarms, real-time surveillance, search-and-rescue, and vehicle-to-vehicle (V2V) communication. In these settings, achieving good performance requires timely delivery of status updates from sources to destinations.

Age of Information (AoI) is a metric that captures timeliness of received information at a destination \cite{kaul2012real,kam2013age,yin17_tit_update_or_wait}. Unlike packet delay, AoI measures the lag in obtaining information at a destination node, and is therefore suited for applications involving time sensitive updates. Age of information, at a destination, is defined as the time that has elapsed since the last received information update was generated at the source. AoI, upon reception of a new update, drops to the time elapsed since generation of the update, and grows linearly otherwise. Over the past decade, there has been a rapidly growing body of work on analyzing AoI for queuing systems \cite{kaul2012real, kam2013age, yin17_tit_update_or_wait,huang2015optimizing,inoue2018general,bedewy2019minimizing}, using AoI as a metric for scheduling policies in networks \cite{kadota2018scheduling,kadota2018scheduling2,talak2018optimizing,tripathi2017age,farazi2018age,tripathi2019whittle} and for monitoring or controlling systems over networks \cite{sun2017remote,sun2019sampling,ornee2019sampling,champati2019performance,klugel2019aoi}. For detailed surveys of AoI literature see \cite{kosta2017age} and \cite{sun2019age_book}.

Typically, AoI is used as a metric for measuring freshness of information being delivered about a source to a monitoring station. It represents a measure of distortion between the state of the system that is expected at the monitor based on past updates and the actual current state of the system. Thus, a larger age corresponds to the monitor having a higher uncertainty about the current state of the system being observed. This, in turn, means that ensuring a low average AoI can lead to higher monitoring accuracy or better control performance. While AoI is a proxy for measuring the cost of having out-of-date information, it may not properly reflect the impact of stale information on system performance. For example, a source might have a high AoI but the monitor might have a good estimate of its state because another source monitoring phenomena nearby sent updates very recently.
 
A typical problem formulation used to model networked monitoring and control settings is one involving multiple sources sending status updates to a central monitoring station. Many prior works have looked at optimizing information freshness metrics for such models under different assumptions on the interference constraints \cite{kadota2018scheduling, talak2018optimizing}, arrival processes \cite{kadota2019minimizing, li2019general}, costs of AoI \cite{kadota2018scheduling2,tripathi2019whittle}, and update sizes \cite{li2019general, tripathi2021computation}. However, all of these works assume that the information across different sources is \textit{decoupled} or \textit{uncorrelated}, i.e. update deliveries from one source only influence the AoI evolution for that source.

This is not strictly true in practice. Many monitoring and control applications involve observing information from correlated or coupled sources. Examples include: vehicular networks where vehicles communicate with their neighbors, multi-agent robotics tasks such as mapping where robots sense overlapping information, and wireless sensor networks where sensors collect spatially correlated updates or exchange information locally.

In such applications, updates from one source often contain information about the current state of other sources. The focus of this work is to understand to role of correlation in designing scheduling policies for information freshness in wireless networks. In Sec.~\ref{sec:model}, we formulate a simple model to analyze weighted-sum average AoI in the presence of correlated sources under wireless interference constraints. In Sec.~\ref{sec:policies}, we use this model to design scheduling policies that can utilize the correlation structure between sources. We formulate a convex problem that solves for the optimal stationary randomized policy and show that it is factor-2 optimal in general. We then develop a Lyapunov drift-based max-weight policy that works well in practice and show that it is also constant factor optimal. In Sec.~\ref{sec:scaling}, we provide scaling results that allow us to understand how the degree of correlation affects information freshness. In Sec.~\ref{sec:robust}, we discuss some alternate ways to model correlation and show that the average AoI for these models remains the same as our proposed model under randomized policies. This highlights the robustness of our results to the way in which correlation is modeled. In Sec.~\ref{sec:learning}, we consider the setting where correlation parameters are unknown and possibly time-varying. Here, we propose a heuristic algorithm called \textit{EMA-max-weight} based on exponential moving averages. This algorithm attempts to both keep track of the correlation parameters and adjust the scheduling decisions in an online manner so as to keep information fresh at the base station. Finally, in Sec.~\ref{sec:sim}, we show numerically that our proposed policies outperform scheduling schemes that ignore the correlation structure inherent in the problem and verify our theoretical results.
 
There has been some prior work in trying to understand how correlation influences information freshness. In \cite{poojary2017real}, the authors consider updates from a single source that are temporally correlated. In \cite{he2018minimizing, he2019joint}, the authors consider a network of cameras with overlapping fields of view and formulate a joint optimization problem that looks at processing and scheduling. In \cite{zhou2020age}, the authors consider a setting with multiple sensors partially monitoring a single source, where updates from at least $M$ sensors are required to reconstruct the state of the source. In \cite{jiang2019status}, the authors consider spatially correlated updates from a random field and study the optimal density at which to place sensors. In \cite{kalor2019minimizing}, the authors consider a two hop setting where sources can send updates to multiple sensors.
\begin{figure}
	\centering
	\includegraphics[width=0.99\linewidth]{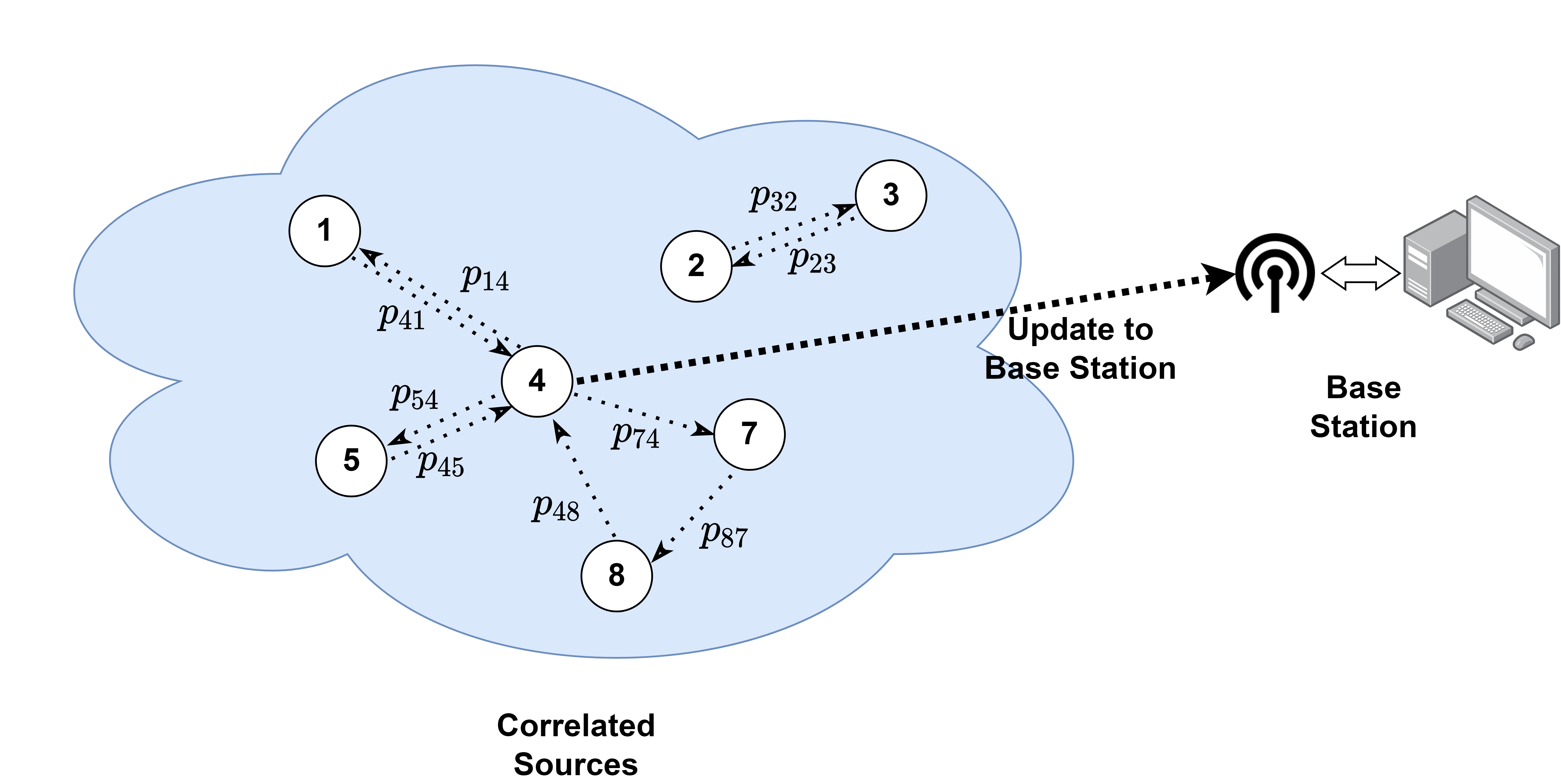}
	\caption{Sources share information locally and send updates to a base station.}
	\label{fig:model}
	\vspace{-1mm}
\end{figure}

To the best of our knowledge, our work is the first one to consider coupled AoI evolution. In our model, the AoI of other sources can drop whenever a source that is correlated with them transmits, since a correlated update can reduce uncertainty about the state of other sources. This couples the AoI evolution of sources and leads to fundamentally new scheduling design and scaling results. 



\section{System Model}
\label{sec:model}

Consider $N$ sources monitoring phenomena of interest and sending updates to a base station. We assume discrete slotted time and assume that only one source can transmit to the base station successfully in any given time-slot. 

We consider a simple model for the correlation structure between sources. At the beginning of every time-slot, each source $i$ collects information about its own state. In addition, with probability $p_{ij}$, the update collected by source $i$ also contains information about the current state of source $j$, for example due to overlapping fields of view between source $i$ and $j$ or due to spatial correlation between the processes being monitored. We assume that this information sharing or overlap happens \textit{independently} across each \textit{ordered pair} of sources and over time. Clearly, a value of $p_{ij} = 0$ suggests that there is never any information at $i$ about $j$, while a value of $p_{ij} = 1$ suggests that $i$ has complete information about $j$ at all times. The overall correlation structure between the sources is described by a matrix $\mathbf{P}$ which contains the pairwise {correlation probabilities}. Figure~\ref{fig:model} depicts an example of such a system, where sources have access to some local information about their neighbors and send updates to a base station. Note that $p_{ii} = 1$ for all sources, since each source is assumed to have information about itself. However, our model can also capture situations where a source fails to obtain information about itself occasionally, by setting $p_{ii} < 1$.

When a particular source transmits to the base station, it sends information about its own state to the base station. However, this update will also contain shared information about some of its neighboring sources. Thus, updates sent to the base station are \textit{correlated}: they can contain information about multiple sources at a time. For example, when source $4$ transmits to the base station in the setup depicted in \autoref{fig:model}, it transmission will contain information not just about itself but also about source $1$ with probability $p_{41}$, source $5$ with probability $p_{45}$ and source $8$ with probability $p_{48}$. 

While we focus on the broadcast interference setting with reliable channels to develop our results and insights, our work can be easily extended to general interference constraints and unreliable channels.

\subsection{Age of Information}
Let $u_i(t)$ be an indicator variable that denotes whether source $i$ transmits to the base station in time-slot $t$. Further let $X_{ij}(t)$ be an indicator variable that denotes whether the current update at source $i$ contains common information about source $j$ in time-slot $t$. From our discussion above, we know that $X_{ij}(t) \sim Bern(p_{ij})$ independent across pairs $(i,j)$ and also over time.

Given this correlation structure, the Age of Information for source $i$ at the base station evolves as follows:
\begin{equation}
\label{eq:AoIEvolution}
A_{i}(t+1) =
\begin{cases}
1, &\text{ if } \sum_{j = 1}^{N} u_j(t) X_{ji}(t) = 1 \\
A_{i}(t)+1, &\text{ otherwise.}
\end{cases}
\end{equation}

\begin{figure}
	\centering
	\includegraphics[width=0.99\linewidth]{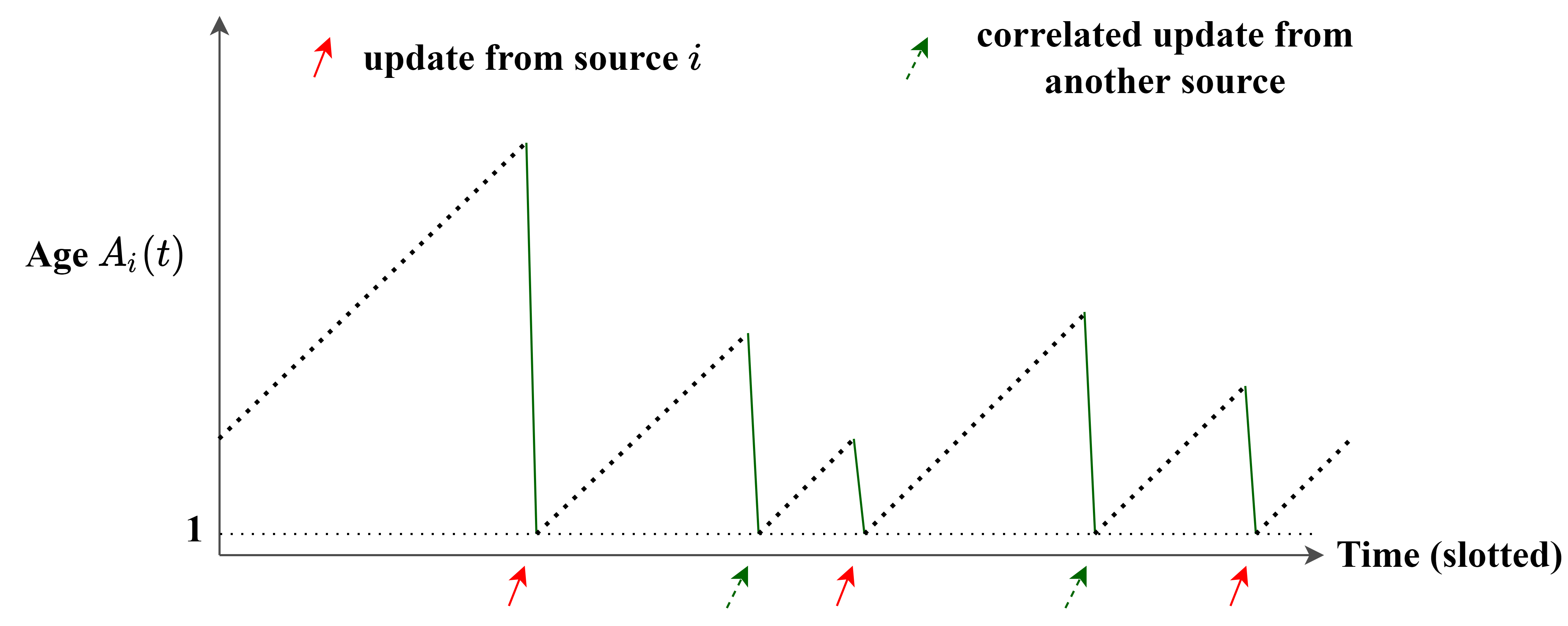}
	\caption{Correlated AoI evolution.}
	\label{fig:AoI}
	\vspace{-1mm}
\end{figure}
The equation \eqref{eq:AoIEvolution} implies that the AoI of source $i$ at the base station drops to $1$ whenever $i$ itself transmits a new update or $j$ transmits a new update containing information about $i$. \autoref{fig:AoI} depicts this AoI evolution. Note that \eqref{eq:AoIEvolution} also assumes that sources only transmit their freshest update, i.e. an older update containing more information about neighbors might not be sent since it was replaced by a newer update with lesser information about neighbors in the next time-slot.

The metric of interest in this work will be average AoI, which is simply the long-term time-average of the AoI process. Specifically,
\begin{equation}
\label{eq:avg_AoI_def}
    \bar{A}_i \triangleq \limsup_{T \rightarrow \infty} \frac{1}{T}  \sum_{t=1}^{T} A_i(t) .
\end{equation}
\subsection{Goal}
Given the probabilistic correlation structure and the AoI evolution described above, we want to design a wireless scheduling policy that minimizes the weighted sum of average AoI across all sources:
\begin{equation}
\label{eq:age_opt_prob}
\underset{\pi}{\operatorname{argmin}} \bigg( \limsup_{T \rightarrow \infty}   \bigg[\frac{1}{T} \sum_{t = 1}^{T} \sum_{i=1}^{N} w_i A_{i}(t) \bigg] \bigg).
\end{equation}

Here, the weights $\{w_1, w_2, ..., w_N\}$ are positive real numbers that denote the relative importance of each source to the overall monitoring or control application.


\section{Scheduling Policies}
\label{sec:policies}
To solve the optimization problem \eqref{eq:age_opt_prob}, we first study stationary randomized policies in Sec.~\ref{sec:srp}. These policies are amenable to analysis and provide key structural insights. Then, in Sec.~\ref{sec:mw}, we develop a Lyapunov drift-based max-weight policy and prove performance bounds for it using the structure of the optimal stationary randomized policy.

\subsection{Stationary Randomized Policies}
\label{sec:srp}

A stationary randomized policy is described by a probability distribution $\bm{\pi}$ over the set of sources, where $\pi_i$ denotes the probability of choosing source $i$. In every time-slot, the policy chooses which source gets to transmit by sampling from the distribution $\pi$ and scheduling decisions are sampled independently across time-slots.

The following theorem relates the average AoI to the scheduling distribution $\bm{\pi}$.

\begin{theorem}
\label{thm:avg_AoI}
Consider any stationary randomized policy with scheduling probabilities $\bm{\pi}$. If $\sum_{j=1}^{N} \pi_j p_{ji} > 0$ for a source $i$, then the average AoI for this source is given by:
\begin{equation}
\label{eq:avg_AoI}
    \bar{A}_i \triangleq \lim_{T \rightarrow \infty} \frac{1}{T} \sum_{t=1}^{T} A_i(t) 
= \frac{1}{\sum_{j=1}^{N} \pi_j p_{ji}}.
\end{equation}
If $\sum_{j=1}^{N} \pi_j p_{ji} = 0$ for some source $i$, then the base station never receives any information regarding this source and its average AoI $\bar{A}_i$ is unbounded.
\end{theorem}
\begin{proof}
See Appendix~\ref{pf:sr_avg_AoI}.
\end{proof}
Using the above theorem, we can 
formulate an optimization problem to find optimal stationary randomized policies. 
\begin{lemma}
Consider the space of stationary randomized policies $\Pi^{sr}$. Finding a policy $\bm{\pi}^* \in \Pi^{sr}$ that minimizes the weighted sum of average AoI is equivalent to solving the following optimization problem: 
\begin{align}
\begin{aligned}
    \label{eq:age_opt_prob_sr}
\underset{\bm{\pi}}{\operatorname{argmin}} & \sum_{i=1}^{N} \frac{w_i}{\sum_{j=1}^{N} \pi_j p_{ji}} \\
\text{s.t. }& \sum_{i=1}^{N} \pi_i \leq 1,\\
& \pi_i \geq 0, \forall i \in [N].
\end{aligned}
\end{align}
\end{lemma}
\begin{proof}
Using the definition of average AoI \eqref{eq:avg_AoI_def} and the expression derived in \autoref{thm:avg_AoI}, we can simplify \eqref{eq:age_opt_prob} to obtain \eqref{eq:age_opt_prob_sr}. The constraints simply represent a valid stationary randomized policy given the assumption that only one source can transmit in any given time-slot.
\end{proof}

Next, we discuss how to solve the optimization problem \eqref{eq:age_opt_prob_sr}. 

\begin{theorem}
\label{thm:sr_struc}
The optimization problem \eqref{eq:age_opt_prob_sr} is convex in the probability distribution $\bm{\pi}$. Further, if the optimal solution $\bm{\pi}^{*}$ to \eqref{eq:age_opt_prob_sr} is such that $\pi^{*}_i > 0, \forall i \in S$ where $S \subseteq [N]$ is a subset of sources, then $\bm{\pi}^{*}$ can be found by solving the following system of nonlinear equations:
\begin{align}
    \sum_{j=1}^{N} p_{ij} w_j \bar{A}^2_j  = \lambda &, \forall i \in S, \text{ and }\\
    \sum_{i \in S} \pi^{*}_i = 1.
\end{align}
Here, $\lambda>0$ is a constant and $\bar{A}_i$ denotes the average AoI for source $i$ under the policy $\bm{\pi}^{*}$ computed using \eqref{eq:avg_AoI}.
\end{theorem}
\begin{proof}
See Appendix~\ref{pf:sr_struc}.
\end{proof}

\autoref{thm:sr_struc} establishes two key results. First, since the optimization problem \eqref{eq:age_opt_prob_sr} is convex, it can be solved efficiently by using a standard solver such as cvx \cite{grant2014cvx}. Second, if the optimal policy involves scheduling some subset of sources a positive fraction of the time, then the quantity $\sum_{j=1}^{N} p_{ij} w_j \bar{A}^2_j$ is constant across all sources in this subset. This can be contrasted with the equivalent result in the uncorrelated case, where the quantity $w_j \bar{A}^2_j$ is constant across all sources \cite{kadota2018scheduling}.
We also recover the well known result from the uncorrelated case, i.e. $\pi^*_i \propto \sqrt{w_i}$ if we set $\mathbf{P} = \mathbf{I}$.

Until now, we have only discussed optimization within the space of stationary randomized policies. The following theorem shows that this class is not too restrictive, i.e. the best stationary randomized policies are at-most a factor of two away from the  best possible scheduling policy \textit{in general}.

\begin{theorem}
\label{thm:sr_2opt}
Consider an optimal stationary randomized policy $\bm{\pi}^{*}$ that is a solution to \eqref{eq:age_opt_prob_sr} and an optimal policy $\bm{\pi}^{opt}$ that solves the general problem \eqref{eq:age_opt_prob}. Let $\mathbf{\bar{A}}^{*}$ denote the average AoIs under $\bm{\pi}^{*}$ and $\mathbf{\bar{A}}^{opt}$ denote the average AoIs under $\bm{\pi}^{opt}$. Then,
\begin{equation}
    \frac{\sum_{i=1}^N w_i \bar{A}^{*}_i}{\sum_{i=1}^N w_i \bar{A}^{opt}_i} \leq 2.
\end{equation}
\end{theorem}
\begin{proof}
See Appendix~\ref{pf:sr_2opt}.
\end{proof}

\subsection{Max-Weight Policy}
\label{sec:mw}
Motivated by the Lyapunov drift-based policies proposed in \cite{kadota2018scheduling,  talak2018optimizing, sun2019age_book}, we next look at an alternative way to design scheduling policies that take correlation into account. Consider the quadratic Lyapunov function given by:
\begin{equation}
\label{eq:lyap2}
    L(t) \triangleq \sum_{i=1}^{N} w_i A^2_i(t).
\end{equation}
Then, the \textit{quadratic max-weight policy} chooses a scheduling decision that minimizes the one-slot Lyapuonv drift in every time-slot. 
\begin{equation}
    \pi^{qmw}(t) = \underset{i \in [N]}{\operatorname{argmin}}~~\mathbb{E} \bigg[ L(t+1) - L(t) \bigg| \mathbf{A}(t) \bigg].
\end{equation}
This simplifies to:
\begin{equation}
\label{eq:qmw}
    \pi^{qmw}(t) = \underset{i \in [N]}{\operatorname{argmin}}~~ \sum_{j = 1}^{N} w_j p_{ij}  A_j(t) \bigg(A_j(t) + 2 \bigg).
\end{equation}

The following theorem provides an upper-bound on the performance of the quadratic max-weight policy.
\begin{theorem}
\label{thm:qmw}
Consider the quadratic max-weight policy $\bm{\pi}^{qmw}$ and an optimal policy $\bm{\pi}^{opt}$ that solves the general problem \eqref{eq:age_opt_prob}. Let $\mathbf{\bar{A}}^{qmw}$ denote the average AoIs under $\bm{\pi}^{qmw}$ and $\mathbf{\bar{A}}^{opt}$ denote the average AoIs under $\bm{\pi}^{opt}$. Then,
\begin{equation}
    \frac{\sum_{i=1}^N w_i \bar{A}^{qmw}_i}{\sum_{i=1}^N w_i \bar{A}^{opt}_i} \leq 4.
\end{equation}
\end{theorem}
\begin{proof}
See Appendix~\ref{pf:qmw}.
\end{proof}

Observe that computing the decisions in the equation above does not require us to explicitly solve for the optimal stationary randomized policy $\bm{\pi}^{*}$. Next, we will develop a max-weight policy that utilizes the optimal stationary randomized policy $\bm{\pi}^{*}$ to get even better performance guarantees. 

Consider an optimal stationary randomized policy $\bm{\pi}^{*}$ that solves the optimization problem \eqref{eq:age_opt_prob_sr}. Using $\bm{\pi}^{*}$, we define the quantities:
\begin{equation}
    \alpha_i \triangleq \frac{w_i}{\sum_{j=1}^{N} \pi^{*}_j p_{ji}}, \forall i \in [N].
\end{equation}
We use the quantities $\alpha_i$ to construct the following linear Lyapunov function:
\begin{equation}
\label{eq:lyapunov}
    L(t) \triangleq \sum_{i=1}^{N} \alpha_i A_i(t).
\end{equation}
Then, our new max-weight policy chooses a scheduling decision that minimizes the one-slot Lyapuonv drift in every time-slot. 
\begin{equation}
\label{eq:mw1}
    \pi^{mw}(t) = \underset{i \in [N]}{\operatorname{argmin}}~~\mathbb{E} \bigg[ L(t+1) - L(t) \bigg| \mathbf{A}(t) \bigg].
\end{equation}

The structure of the max-weight policy can be obtained by simplifying the expression in \eqref{eq:mw1} above:
\begin{align}
    \label{eq:mw2}
    \begin{aligned}
        \pi^{mw}(t) &= \underset{i \in [N]}{\operatorname{argmin}}\bigg(\sum_{j = 1}^{N} p_{ij} \alpha_j A_j(t)\bigg) \\
        &= \underset{i \in [N]}{\operatorname{argmin}} \bigg(\alpha_i A_i(t) + \sum_{j \neq i} p_{ij} \alpha_j A_j(t)\bigg).
    \end{aligned}
\end{align}
Note that the max-weight policy proposed in \cite[Sec.\ 3.2.4]{sun2019age_book} for the uncorrelated setting schedules the source $i$ with the highest value of $\sqrt{w_i} A_i(t)$. On the other hand, our correlated max-weight policy \eqref{eq:mw2} adds up the ``value'' of each possible AoI reduction including correlated updates, weighted by the of probability of correlation. 

The following theorem shows that the new max-weight policy defined in \eqref{eq:mw2} enjoys a similar factor-2 optimality guarantee as the optimal stationary randomized policy which is better than the factor-4 guarantee for the quadratic max-weight policy described in \eqref{eq:qmw}.
\begin{theorem}
\label{thm:mw_4opt}
Consider the max-weight policy $\bm{\pi}^{mw}$ and an optimal policy $\bm{\pi}^{opt}$ that solves the general problem \eqref{eq:age_opt_prob}. Let $\mathbf{\bar{A}}^{mw}$ denote the average AoIs under $\bm{\pi}^{mw}$ and $\mathbf{\bar{A}}^{opt}$ denote the average AoIs under $\bm{\pi}^{opt}$. Then,
\begin{equation}
    \frac{\sum_{i=1}^N w_i \bar{A}^{mw}_i}{\sum_{i=1}^N w_i \bar{A}^{opt}_i} \leq 2.
\end{equation}
\end{theorem}
\begin{proof}
See Appendix~\ref{pf:mw_4opt}.
\end{proof}

While we show a factor of two optimality result above, we will show via simulations in Sec.~\ref{sec:sim} that the max-weight policy performs almost as well as the theoretical lower bound derived in Appendix \ref{pf:sr_2opt} and also outperforms the optimal stationary randomized policy in practice.

\section{Scaling}
\label{sec:scaling}
In this section, we consider how correlation improves information freshness as network sizes scale. For deriving our scaling results, we will focus on the equal weights setting, i.e. $w_i = \frac{1}{N}, \forall i \in [N]$. However, similar scaling results will hold qualitatively for general weight configurations.

The lemma below characterizes the minimum average AoI of a network with $N$ {uncorrelated} sources with equal weights. We will use this as a comparison baseline to see how the degree of correlation improves AoI.
\begin{lemma}
\label{lem:uncorr}
Consider $N$ \textit{uncorrelated} sources sending updates to a base station. Let the scheduling weights for all sources be equal, i.e. $w_i = \frac{1}{N}, \forall i$. Then, the optimal weighted sum AoI satisfies the following:
\begin{equation}
    \sum_{i=1}^{N} w_i \bar{A}^{opt}_i = \frac{N+1}{2} \sim \Theta(N).
\end{equation}
\end{lemma}
\begin{proof}
In the setting with symmetric weights, the greedy or the round-robin policy is known to be optimal \cite{kadota2018scheduling}. It is easy to see that the average AoI under the round-robin policy is simply the average of the sequence $1,2,..., N$ which is $\frac{N+1}{2}$.
\end{proof}

\subsection{An Upper Bound}
We derive a general upper bound for the weighted-sum average AoI by representing information about the correlation matrix $\mathbf{P}$ for $N$ sources with a directed graph.

Consider a correlation threshold $p \in (0,1)$. We are interested in the entries of the correlation matrix above this threshold, i.e the pairs of sources that are significantly correlated. To do so, we construct a directed graph $\mathcal{G}(V,E)$ on the set of sources. For every ordered pair of sources $(i,j)$ such that $p_{ij} > p$, we add the edge $(i,j)$ to the graph $\mathcal{G}$. Trivially, every node in $\mathcal{G}$ must have a self-loop, since $p_{ii} = 1, \forall i$. We will show that the average AoI of the network can be upper-bounded by analyzing the properties of these constructed graphs. To derive this upper bound,
we first need to define the notion of a vertex cover for a directed graph.


\textbf{Vertex Cover}: Given a directed graph $\mathcal{G}(V,E)$, a vertex cover is defined to be a set of vertices $S \subseteq V$ if for every vertex $i \in V$, there exists a vertex $j \in S$ such that the edge $(j,i)$ is in the set $E$.

The following theorem relates the average AoI to the size of the \textit{minimum vertex cover} of the graph $\mathcal{G}$ constructed using the correlation threshold $p$.

\begin{theorem}
\label{thm:scaling}
Consider $N$ sources with the correlation matrix $\mathbf{P}$. Given a correlation threshold $p>0$, construct a directed graph $\mathcal{G}$ that represents pairs of source with correlation higher than the threshold. Assuming equal weights $w_i =\frac{1}{N}, \forall i$, the optimal weighted sum average AoI satisfies:
\begin{equation}
    \sum_{i=1}^{N} w_i \bar{A}^{opt}_i \leq \frac{N_{cov}}{p},
\end{equation}
where $ N_{cov}$ is the size of a minimum vertex cover for the graph $\mathcal{G}$ and $p$ is the correlation threshold. 
\end{theorem}
\begin{proof}
See Appendix~\ref{pf:scaling}.
\end{proof}

This upper-bound allows us to relate the degree of correlation or information sharing between the sources to the average AoI. If a small subset of sources, of say size $O(\log N)$, are highly correlated with all other sources in the network, then average AoI with correlation is only $O(\log N)$ as well. In this case, correlation leads to a significant reduction in the average AoI compared to the uncorrelated case, which is $O(N)$ as shown in Lemma~\ref{lem:uncorr}. If no such small vertex coverings exist or if correlation probabilities are very small, then a scheduler likely needs to communicate with a large fraction of all the sources and the average AoI would grow as $O(N)$, similar to the uncorrelated case. 

Further, we show that it is possible to construct a correlation matrix such that the average AoI of the correlated max-weight given by \eqref{eq:mw2} is $O(1)$ while the average AoI of the uncorrelated max-weight policy from \cite{kadota2018scheduling} is $\Theta(N)$. This means that the performance gap between policies that consider and ignore correlation can grow linearly with the size of the network.
\begin{theorem}
\label{lem:compare}
Given any number of sources $N$ with equal weights $\frac{1}{N}$, we consider two policies - the correlated max-weight policy $\pi^{mw}$ proposed in Sec.~\ref{sec:mw} and the max-weight policy that does not take correlation into account $\pi^{u}$ which was proposed in \cite{kadota2018scheduling}. Let the average AoI of a source $i$ under $\pi^{mw}$ be $\bar{A}^{mw}_i$ and under $\pi^{u}$ be $\bar{A}^{u}_i$. Then, there exists a correlation matrix $\mathbf{P} \in [0,1]^{N \times N}$ such that the following holds:
\begin{equation}
    \frac{\sum_{i=1}^{N} w_i \bar{A}^{u}_i }{\sum_{i=1}^{N} w_i \bar{A}^{mw}_i } \sim \Omega(N).
\end{equation}
\end{theorem}
\begin{proof}
Consider $N$ sources with the following correlation matrix:
\begin{equation}
    \mathbf{P} = \begin{bmatrix}
    1 &p &p &\cdots &p\\
    p &1 &0 &\cdots &0\\
    p &0 &1 &\cdots &0\\
    \vdots &\vdots &\cdots &1 &\vdots\\
    p &0 &0 &\cdots &1\\
    \end{bmatrix}.
\end{equation}
The stationary randomized policy $\bm{\pi}^{(1)}$ that puts all scheduling weight on source $1$ achieves a weighted-sum AoI of:
\begin{equation}
    \sum_{i=1}^{N} w_i \bar{A}_i = \frac{1}{N} + \sum_{i=2}^{N} \frac{1}{N} \frac{1}{p} \leq \frac{1}{p}.
\end{equation}
Thus, the optimal stationary randomized policy $\bm{\pi}^{*}$ also achieves weighted-sum average AoI that is at-least as good as that of $\bm{\pi}^{(1)}$. This implies:
\begin{equation}
    \sum_{i=1}^{N} w_i \bar{A}^{*}_i \leq \frac{1}{p},
\end{equation}
where $\bar{A}^{*}_i$ is the average AoI of source $i$ under the optimal stationary randomized policy $\bm{\pi}^{*}$. In Appendix~\ref{pf:mw_4opt}, we show that the performance of the max-weight policy is upper-bounded by the policy of the optimal stationary randomized policy. Thus, we get:
\begin{equation}
\label{eq:cex1}
    \sum_{i=1}^{N} w_i \bar{A}^{mw}_i \leq \sum_{i=1}^{N} w_i \bar{A}^{*}_i \leq \frac{1}{p},
\end{equation}
where $\bar{A}^{mw}_i$ is the average AoI of source $i$ under the max-weight policy $\bm{\pi}^{mw}$ described by \eqref{eq:mw2}. 

Now consider the performance of the uncorrelated max-weight policy $\bm{\pi}^{u}$. Since all the weights are symmetric, the optimal policy is greedy or max-AoI-first. In Appendix~\ref{pf:rr_maf}, we show that the max-AoI-first policy behaves similar to a round-robin policy on sources $2,...,N$ while occasionally scheduling source $1$. Specifically, we derive a lower bound on the weighted-sum AoI under policy $\bm{\pi}^{u}$:
\begin{equation}
\label{eq:cexi}
    \sum_{i=1}^{N} w_i \bar{A}^{u}_i \geq \frac{(N-1)^2}{2N - (N+1)(1-p)^{N-1}}.
\end{equation}
This bound holds under the assumtion that $p \geq \frac{1}{N-1}$. Combining \eqref{eq:cex1} and \eqref{eq:cexi}, we get:
\begin{equation}
    \frac{\sum_{i=1}^{N} w_i \bar{A}^{u}_i }{\sum_{i=1}^{N} w_i \bar{A}^{mw}_i } \geq  \frac{p(N-1)^2}{2N - (N+1)(1-p)^{N-1}}.
\end{equation}
Assuming $p$ to be a fixed constant that does not depend on $N$ completes the proof. Note that a similar result also holds for $p$ scaling  with $N$, for example if $p \sim \frac{1}{N^\beta}, \beta < 1$ the performance gap is $\Omega(N^{1-\beta})$.
\end{proof}
The intuition behind this result is straightforward: if there is one source that is correlated even by a small amount with all other sources then scheduling just that source all the time should be sufficient. The gap in performance is achieved by assuming that other sources are not correlated with each other. 

Theorem~\ref{lem:compare} provides a key insight: \textit{the gain in performance cannot be obtained by using policies that ignore correlation}. It is necessary to design scheduling policies that take the correlation structure into account, especially for correlation graphs where the degree distribution is highly skewed. Next, we consider scaling in the special case where the correlation matrix can be represented by random geometric graphs. These graphs are commonly used to model wireless sensors networks monitoring spatial phenomenon.
\subsection{Random Geometric Graphs}
Consider $N$ sources distributed uniformly at random on the unit square $[0,1]^{2}$. Each source has information about itself, by definition. Thus $p_{ii} = 1, \forall i$. Further, if the distance between sources $i$ and $j$ is less than a threshold $r$, then we set $p_{ij} = p$ and $p_{ji} = p$, otherwise we set $p_{ij} = 0$ and $p_{ji} = 0$. This leads to a symmetric correlation matrix $\mathbf{P}$. Constructing a graph that connects sources with correlation leads to the random geometric graph $\mathcal{G}(N,r)$.

The following theorem looks at the scaling of AoI under this geometric correlation structure.

\begin{theorem}
\label{thm:rgg}
Consider a symmetric correlation matrix generated by creating a random geometric graph $\mathcal{G}(N,r)$ on the two dimensional unit square and setting correlation probabilities for neighbors to be $p$. Assuming equal weights $w_i = \frac{1}{N}, \forall i$,  the weighted sum average AoI satisfies:
\begin{equation}
     \sum_{i=1}^{N} w_i \bar{A}^{opt}_i  \leq
    \frac{2}{p r^2}.
\end{equation}
\end{theorem}
\begin{proof}
See Appendix~\ref{pf:rgg}.
\end{proof}

Note that the connectivity threshold of a random geometric graphs occurs at $r \sim \Theta\bigg(\sqrt{\frac{\log N}{N}} \bigg)$. For this choice of $r$, we can see that the overall AoI of the network is $O\bigg(\frac{N}{p\log N}\bigg)$. This leads to a factor of $\log N$ reduction over the uncorrelated case analyzed in Lemma~\ref{lem:uncorr}, assuming $p$ is a constant. Further, if $r \sim \Theta(1)$, then the AoI is also $\Theta(1)$ and there is an $O(N)$ reduction compared to the uncorrelated case. 


\section{Robustness}
\label{sec:robust}
In this section, we consider a different way to model correlation between a set of sources. We will show that for stationary randomized policies, this model is equivalent to our proposed model. Later, in Sec.~\ref{sec:sim}, we will show via numerical results that the scheduling policies designed in Sec.~\ref{sec:policies} also perform well for this new correlation structure. This suggests that if the ``amount of correlation'' is the same on average, then the AoI performance tends to be similar, regardless of how correlation is modeled.  

As before, consider a set of $N$ sources and a correlation matrix $\mathbf{P}$ with entries $p_{ij} \in [0,1]$ that represent the amount of information shared at source $i$ about source $j$. However, instead of a Bernoulli random variable indicating whether $i$ either does or does not currently have information about $j$, we now assume that $i$ always has a constant $p_{ij}$ fraction of information regarding the state of $j$. In other words, an update from $i$ can reduce the current uncertainty regarding $j$ by a fraction of $p_{ij}$. This correlation structure is convenient for modeling settings where different sources are sensing information and have \textit{partially} overlapping ranges of sensing, for example cameras with overlapping fields of view as studied in \cite{he2018minimizing,he2019joint}. Thus, $p_{ij}$ can be viewed as the fraction of source $j$'s range that source $i$ also covers.

Consequently, the evolution of AoI for this model  differs from the one studied in earlier sections. Let $u_i(t)$ be indicator variables that denote whether source $i$ transmits in time-slot $t$ or not. Then, AoI for source $i$ evolves as follows:
\begin{equation}
\label{eq:AoIEvolution2}
A_{i}(t+1) =
(1-p_{ji}) A_{i}(t) + 1, \text{ if } u_j(t) = 1.
\end{equation}
The equation above formalizes the notion that uncertainty regarding source $i$ drops by a fraction $p_{ji}$, when $j$ sends an update.

The following theorem establishes the equivalence between this new model and the one proposed in Sec.~\ref{sec:model}, under stationary randomized policies. In fact, we prove the equivalence result for a much more general class of correlation structures, where correlation is defined by i.i.d. random variables $X_{ij}(t) \in [0,1]$ such that $\mathbb{E}[X_{ij}(t)] = p_{ij}$. Our original model assumes $X_{ij}(t) \sim Bern(p_{ij})$, while the model proposed in this section assumes $X_{ij}(t) = p_{ij}$ and both of them belong to this class of correlation structures.

\begin{theorem}
\label{thm:equi_sr}
Consider $N$ sources and a correlation matrix $\mathbf{P}$ with the AoIs for each source evolving according to:
\begin{equation}
    A_{i}(t+1) = A_i(t) + 1 - \bigg( \sum_{j=1}^{N} u_j(t) X_{ji}(t) \bigg) A_i(t), \forall i, t.
\end{equation}
Here $X_{ji}(t)$ are i.i.d random variables such that $ X_{ji}(t) \in [0,1]$ and $\mathbb{E}[X_{ji}(t)] = p_{ji}$. Given any stationary randomized policy with scheduling probabilities $\bm{\pi}$, if $\sum_{j=1}^{N} \pi_j p_{ji} > 0$ for a source $i$, then the average AoI for this source is given by:
\begin{equation}
    \bar{A}_i \triangleq \lim_{T \rightarrow \infty} \mathbb{E}\bigg[\frac{1}{T} \sum_{t=1}^{T} A_i(t) \bigg]
= \frac{1}{\sum_{j=1}^{N} \pi_j p_{ji}}.
\end{equation}
If $\sum_{j=1}^{N} \pi_j p_{ji} = 0$ for some source $i$, then the base station never receives any information regarding this source and its average AoI $\bar{A}_i$ is unbounded.
\end{theorem}
\begin{proof}
See Appendix~\ref{pf:equi_sr}.
\end{proof}

Observe that the expression for the average AoI is the same as the one derived in Theorem~\ref{thm:avg_AoI}. Thus the procedure to find the optimal stationary randomized policy and the policy itself also remain the same. We will show later via simulations that the performance of the max-weight policy is also similar for different distributions of $X_{ji}(t)$, which suggests that our analysis is fairly robust to the way in which correlation is modeled. 


\section{Learning The Correlation Matrix}
\label{sec:learning}
Until now, we have focused on cases where the correlation structure is known in advance and fixed over time, and we use this information to analyze and optimize AoI. In this section, we consider the setting when the correlation matrix is unknown and possibly varying with time.




\subsection{Online Setting}
\label{sec:ema}
Learning the correlation matrix in an online setting where the $\mathbf{P}$ changes over time, even slowly, is a challenging problem. This setting is of interest because correlation between source tends to be time-varying in practice, especially in settings involving mobility.

We want to implement a max-weight style policy that gradually updates its policy parameters to be able to track changes in the environment. However, note that the max-weight policy proposed in Sec.~\ref{sec:mw} in \eqref{eq:mw2} requires us to solve for the optimal stationary randomized policy. Thus, whenever the correlation matrix changes, we would have to recalculate $\bm{\pi}^{*}$ and the Lyapunov function weights $\alpha_i$. To avoid this added computation, we will use the quadratic max-weight policy, given by \eqref{eq:qmw}. Recall that this policy is based on a quadratic Lyapunov function and does not require us to calculate the the optimal stationary randomized policy $\bm{\pi}^{*}$ repeatedly as $\bm{P}$ varies.

Algorithm~\ref{alg:EMA-MW} uses an exponential moving average to keep track of the correlation probabilities and then runs the \textit{quadratic} max-weight scheduler from \eqref{eq:qmw} using the estimated correlation matrix. We call this the \textit{EMA-max-weight} policy. 
\begin{algorithm}
	\DontPrintSemicolon
	\SetKwInOut{Input}{Input}\SetKwInOut{Output}{Output}
	\Input{parameter $\alpha > 0$}
	\BlankLine
	Start by assuming no correlation, i.e. set \[\mathbf{\hat{P}}(1) = \mathbf{I}\] \\		
	\While{ $t \in 1,...,T$ }{
		Run quadratic max-weight using $\mathbf{\hat{P}}(t)$: \[ s = \underset{i \in [N]}{\operatorname{argmin}}~~ \sum_{j = 1}^{N} w_j \hat{p}_{ij}(t) A_j(t) \bigg(A_j(t) + 2 \bigg).\] \;
		Schedule source $s$ and receive correlated updates \;
		Update AoIs for every source $j$:
		\[
		A_j(t+1) = \begin{cases}
		1, &\text{ if }s\text{ sent an update about }j,\\
		A_j(t) + 1, &\text{ otherwise.}
		\end{cases}
		\]
		\;
		Update the correlation matrix, for source $s$
		\[
		\hat{p}_{sj}(t+1) = \begin{cases}
		(1-\alpha)\hat{p}_{sj}(t) + \alpha, &\text{ if }s\text{ sent an update about }j,\\
		(1-\alpha)\hat{p}_{sj}(t), &\text{ otherwise.}
		\end{cases}\]
		For sources other than $s$:
		\[
		\hat{p}_{ij}(t+1) = \hat{p}_{ij}(t), \forall i \neq s.
		\]
		\;
	}		
	\caption{Exponential Moving Average Max-Weight}
	\label{alg:EMA-MW}
\end{algorithm}

Intuitively, if the probabilities change slowly over time, the exponential moving average estimate should be able to closely track the actual correlation matrix and the EMA-max-weight policy would perform similar to a max-weight policy that knows the entire sequence of correlation matrices in advance. We confirm that this is indeed the case via simulations in Sec.~\ref{sec:sim}. 

\section{Numerical Results}
\label{sec:sim}
First, we consider random geometric graphs and see how different scheduling policies perform as the number of sources $N$ increases. In particular, we simulate graphs $\mathcal{G}(N,r)$ on the unit square $[0,1]^2$ where $r = 1.1\sqrt{\frac{\log N}{N}}$ is slightly above the connection threshold. For each pair of nodes $(i,j)$ in this graph, we set $P_{ij} = P_{ji} = 0.7$ if the nodes are closer than the distance $r$ and set $P_{ij} = P_{ji} = 0$ otherwise. We fix all weights to be equal, i.e. $w_i = \frac{1}{N}$.

\begin{figure}
	\centering
	\includegraphics[width=0.99\linewidth]{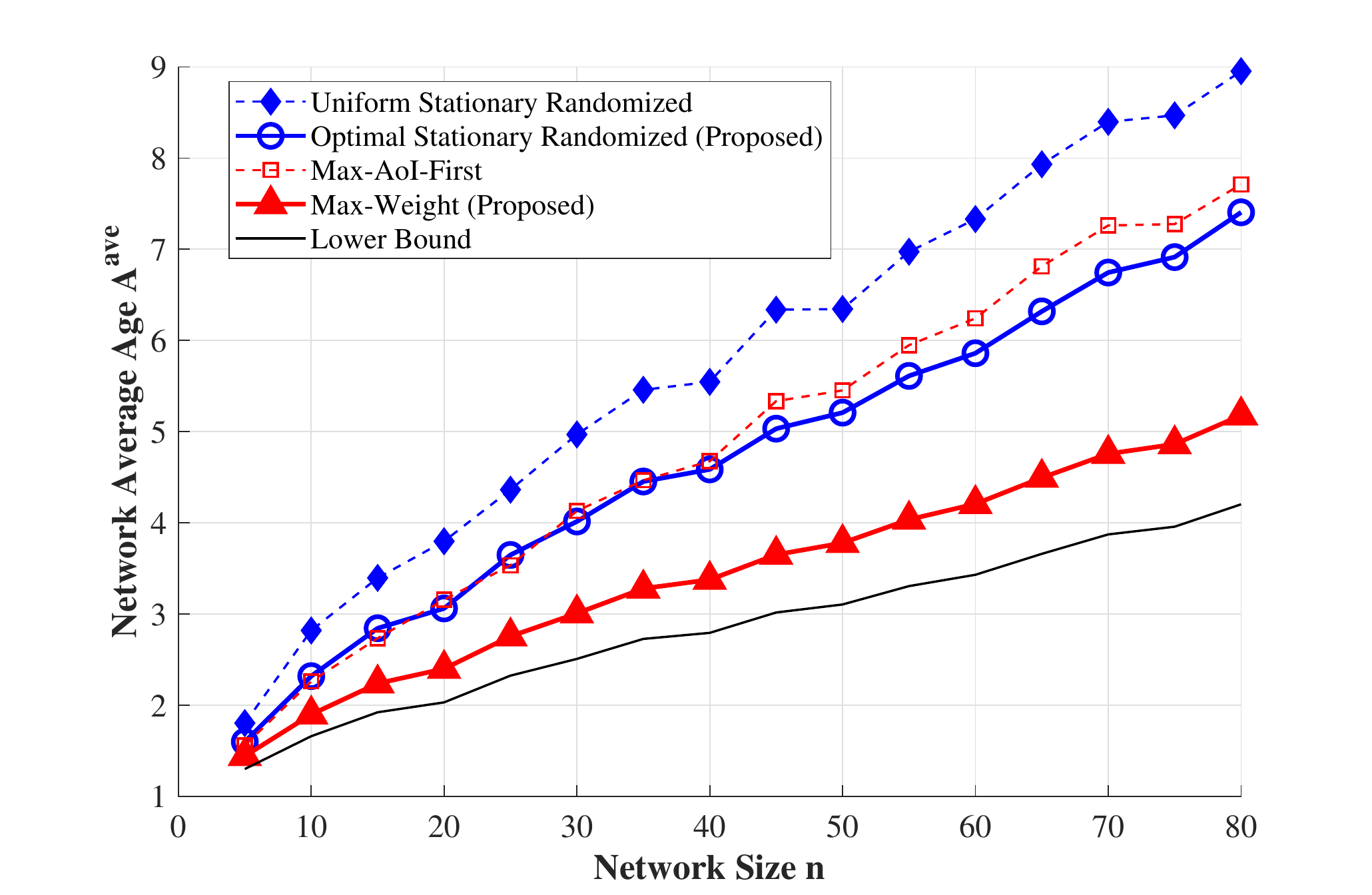}
	\caption{Average AoI vs network size $N$ for random geometric graphs.}
	\label{fig:AoI_rgg}
	\vspace{-1mm}
\end{figure}
For each value of $N$, we compare the performance of four different policies - 1) the uniform stationary randomized policy which would have been the optimal stationary randomized policy if we ignore correlation, 2) our optimal stationary randomized policy which solves \eqref{eq:age_opt_prob_sr}, 3) the max-Age-first or greedy policy which would have been the optimal max-weight policy if we ignore correlation and 4) our proposed max-weight policy from Sec.~\ref{sec:mw}. 

Figure~\ref{fig:AoI_rgg} plots the performance of these four policies as $N$ increases. Each data point is computed by averaging over $10$ random graph instances and running the policy for $10000$ time-slots for each such instance. We also plot a lower bound for average AoI computed using \eqref{eq:lb6} from Appendix~\ref{pf:sr_2opt}. We observe that our proposed methods clearly outperform policies that ignore the correlation between sources for scheduling. In particular, our proposed max-weight policy outperforms max-AoI-first by almost 33\%.

\begin{figure}
	\centering
	\includegraphics[width=0.99\linewidth]{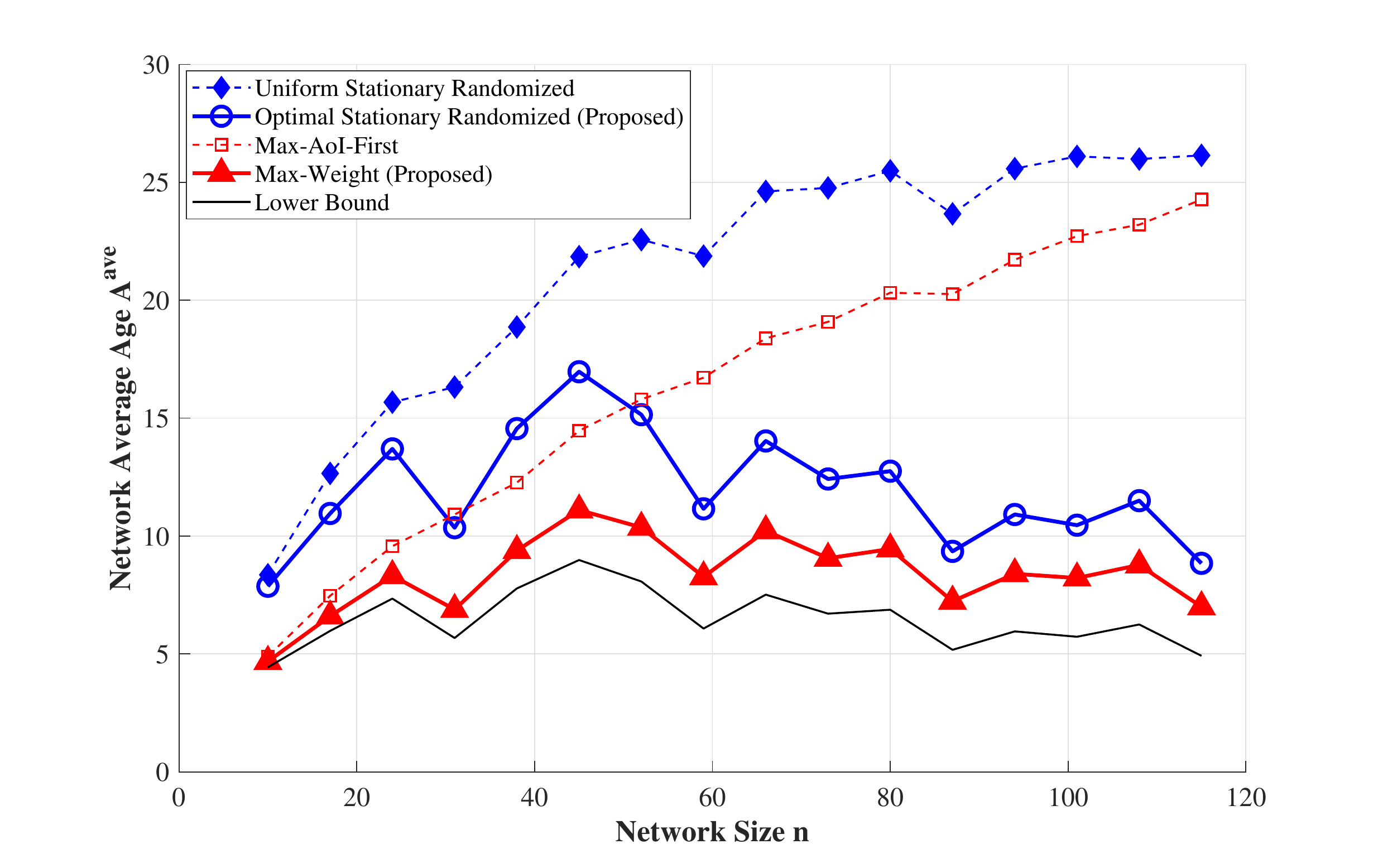}
	\caption{Average AoI vs network size $N$ for hyperbolic geometric graphs.}
	\label{fig:AoI_sf}
	\vspace{-1mm}
\end{figure}
Next, we perform the same exercise for hyperbolic geometric graphs. Hyperbolic geometric graphs are generated by choosing points on a two-dimensional hyperbolic space and connecting vertices that are closer than the distance $R$ where distance is measured along hyperbolic geodesics. The most important feature of hyperbolic random graphs is that when their parameters are chosen appropriately, the degree distribution of nodes becomes \textit{scale-free} \cite{krioukov2010hyperbolic,papadopoulos2010greedy}. Importantly, for our simulation setup, we expect that scale-free degree distributions lead to correlation matrices with lower vertex covering numbers. According to Theorem~\ref{thm:scaling}, this should lead to larger performance gaps between policies that consider correlation and policies that ignore it. Figure~\ref{fig:AoI_sf} confirms this: as the number of sources increase, the gap in performance between our proposed policies and policies that ignore correlation also increases. 

Next, we compare the performance of the optimal stationary randomized policy and our proposed max-weight policy for different correlation models. Specifically, we consider three correlation models:
1) Bernoulli correlation: $X_{ij}(t) \sim Bern(p_{ij})$, which is the main focus of this work, 2) Constant correlation: $X_{ij}(t) = p_{ij}$ as introduced in Sec.~\ref{sec:robust}, and 3) Uniform correlation: $X_{ij}(t) = p_{ij} + Unif\big([-0.1,0.1]\big)$, i.e. correlation is chosen uniformly at random from the interval $[p_{ij} - 0.1, p_{ij}+0.1]$.

\begin{figure}
	\centering
	\includegraphics[width=0.99\linewidth]{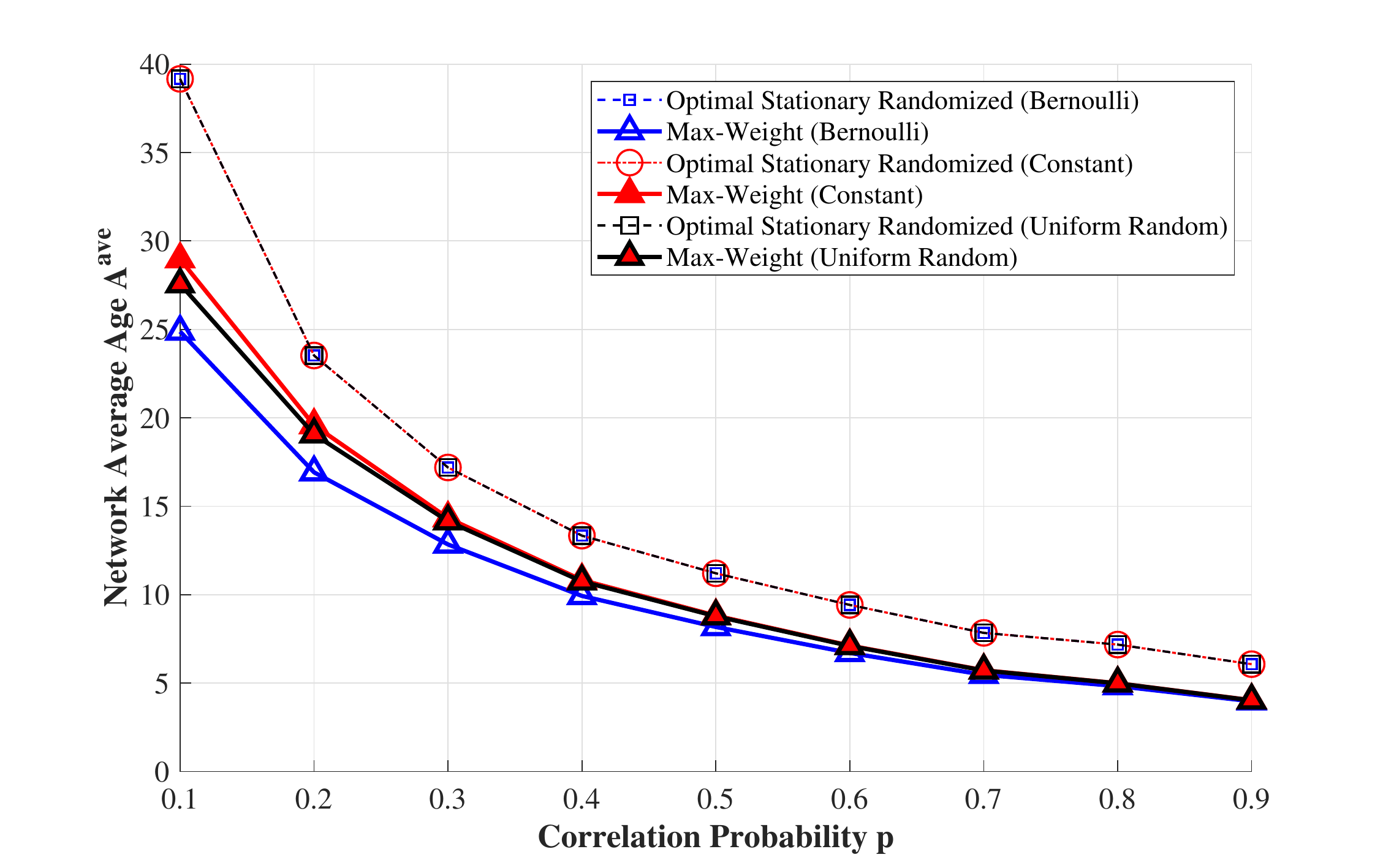}
	\caption{Average AoI vs correlation probability $p$.}
	\label{fig:AoI_corr_models}
	\vspace{-1mm}
\end{figure}

Figure~\ref{fig:AoI_corr_models} compares the performance of the two policies (optimal stationary randomized and max-weight) on random geometric graphs $\mathcal{G}(N=90,r=0.25)$ while varying the correlation parameter $p$ from $0.1$ to $0.9$ for the three different correlation models discussed above. 

We observe that for the stationary randomized policy,  the average AoI values are the same across correlation models for each value of $p$. This is consistent with the result we derived in Theorem~\ref{thm:equi_sr}. For the max-weight policy, the average AoI values are close to one another for each value of $p$. Further, as the correlation increases, the gap between the average AoI for different correlation models decreases. Combined with our observation from Sec.~\ref{sec:robust}, where we showed that average AoI under stationary randomized policies is the same for a large class of correlation models, this supports the claim that our results are fairly robust to the way in which correlation is modeled. Another important observation from Figure~\ref{fig:AoI_corr_models} is the inverse dependence of the average AoI on the correlation probability $p$, consistent with our upper-bound for random-geometric graphs from Theorem~\ref{thm:rgg}.

\begin{figure}
	\centering
	\includegraphics[width=0.99\linewidth]{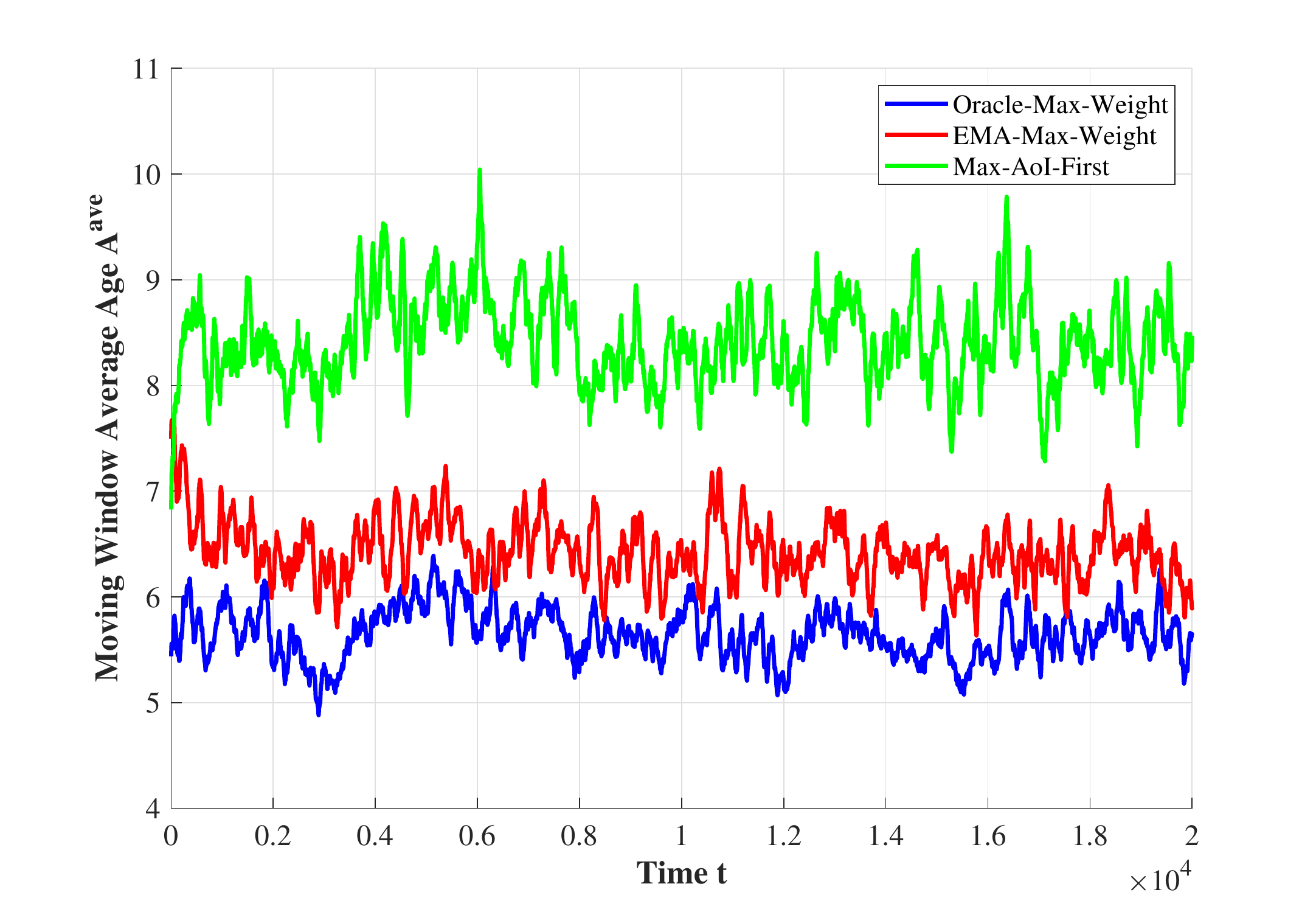}
	\caption{Moving window average AoI vs time.}
	\label{fig:AoI_ema}
	\vspace{-1mm}
\end{figure}
Finally, we consider a setting with time-varying correlation probabilities. Specifically, we consider a random geometric graph with $N=90$ sources and a connectivity radius of 0.25 on the unit square $[0,1]^2$. However, unlike the previous simulations, we assume that these sources are \textit{mobile} and move according to Brownian motion on $[0,1]^2$ with a maximum velocity of $0.01$. As the sources move around randomly, the distances between them change and so does the corresponding correlation matrix $\mathbf{P}$. Figure~\ref{fig:AoI_ema} plots the performance of three different scheduling policies in this time-varying setting. To measure the performance, we consider a windowed time-average of the network AoI with a window size of $100$. As the sources connectivity changes, we see the windowed time-average AoI change in response for each scheduling policy.  

The three policies we consider are as follows. First, we consider the max-AoI-first policy which is completely oblivious of the correlation while making scheduling decisions. Second, we consider the EMA-max-weight policy from Sec.~\ref{sec:ema} which has no information about correlation in the beginning but gradually learns the matrix $\mathbf{P}$ and adapts to changes in it using an exponential moving average. Specifically, we set the learning rate $\alpha = 0.4$. Third, we consider the hypothetical \textit{oracle-max-weight} policy. This is an omniscient policy that knows the current correlation matrix exactly in each time-slot and uses this information to run the quadratic max-weight scheduler \eqref{eq:qmw}.

We observe that EMA-max-weight is able to track the performance of the oracle-max-weight policy with only a small gap, indicating that it is able to learn the correlation structure and adapt in response to it. Further, max-AoI-first is not able to do so and has a larger gap in performance compared to the oracle-max-weight policy.

\section{Conclusion}
In this work, we formulated a simple model to study the timely monitoring of correlated sources over a wireless network. Using this model, we proposed new scheduling policies that optimize weighted-sum average Age of Information (AoI) in the presence of correlation. These policies have constant-factor optimality guarantees. We derived scaling results that illustrate how AoI improves in large networks in the presence of correlation and discussed how our model is relatively robust to correlation modeling assumptions. Lastly, We also developed a novel approach based on exponential moving averages that schedules correlated sources in a time-varying setting.

Important directions of future work involve proving performance bounds on EMA-max-weight under assumptions on how quickly the correlation matrices change, and incorporating more general AoI cost functions. A drawback that is worth mentioning is that we consider correlation to be a pairwise notion - information sharing happens only between pairs of sources. However, modeling more general notions of correlation and coupling between sources while still keeping analysis tractable is a challenging open problem.

\section{Acknowledgements}
This work was supported in part by NSF Grant CNS-1713725.

\bibliographystyle{ACM-Reference-Format}
\bibliography{bibliography_2}

\appendix
\section{Proof of Theorem~\ref{thm:avg_AoI}}
\label{pf:sr_avg_AoI}
Consider a stationary randomized policy with the scheduling probabilities given by $\bm{\pi}$. Then, the probability that the base station receives an update about source $i$ in any time-slot $t$ is given by
\begin{equation}
    \sum_{j=1}^{N} \mathbb{P}\big(j\text{ has information about }i\big)\mathbb{P}\big(j\text{ transmits}\big) = \sum_{j=1}^{N}  p_{ji} \pi_j.
\end{equation}
For simplicity, we will refer to the quantity $\sum_{j=1}^{N}  p_{ji} \pi_j$ as $r_i$. Clearly, if $r_i = 0$, then the base station never receives any information regarding source $i$ and the average AoI $\bar{A}_i$ grows to be unbounded. This proves one part of \autoref{thm:avg_AoI}.

Next, we focus on the case when $r_i > 0$. In each time-slot, the base station receives a new update about source $i$ with probability $r_i$, \textit{independent} of events in other time-slots. Thus, intervals between two consecutive update deliveries from source $i$ to the base station are geometrically distributed i.i.d. random variables with the parameter $r_i$. Let $I_1, I_2, ..., $ be i.i.d. geometric random variables that denote update inter-arrival periods for source $i$. Further, let $K$ be the largest integer such that $\sum_{k=1}^{K} I_k \leq T$, i.e. there are $K$ update deliveries over the first $T$ time-slots. Note that $K$ itself is a random variable. Then, the average AoI of source $i$ is given by:
\begin{align}
\begin{aligned}
    \bar{A}_i &= \limsup_{T \rightarrow \infty} \frac{1}{T}   \sum_{t=1}^{T} A_i(t)  \\
    &= \limsup_{T \rightarrow \infty} \frac{1}{T}  \sum_{k=1}^{K} \sum_{j=1}^{I_k} j  \\
    &= \limsup_{T \rightarrow \infty} \frac{1}{T}  \sum_{k=1}^{K} \frac{I^2_k + I_k}{2}  \\
    &= \frac{\mathbb{E} [I^2_1 + I_1]}{2\mathbb{E} [I_1]}.
\end{aligned}
\end{align}
The last step above holds by applying the elementary renewal-reward theorem and the law of large numbers. $I_1$ is simply a geometric random variable with the parameter $r_i$. Thus, the equation above can be further simplified to:
\begin{align}
    \begin{aligned}
        \bar{A}_i &= \frac{1}{2} + \frac{\mathbb{E} [I^2_1]}{2\mathbb{E} [I_1]}\\
        &= \frac{1}{2} + \frac{(2-r_i)/r^2_i}{2/r_i}\\
        &= \frac{1}{r_i}.
    \end{aligned}
\end{align}

We use the moments of geometric random variables to derive the expression above. Since $r_i = \sum_{j=1}^{N}  p_{ji} \pi_j$, this completes the proof.

\section{Proof of Theorem~\ref{thm:sr_struc}}
\label{pf:sr_struc}
We will first show that the objective function of the optimization problem \eqref{eq:age_opt_prob_sr} is convex in the scheduling probabilities $\bm{\pi}$. To do so, we compute the Hessian of the average AoI of source $i$ (given by $\bar{A}_i$) with respect to $\bm{\pi}$. 
\begin{align}
    \begin{aligned}
        \mathbf{H}(\bar{A}_i) &\triangleq \begin{bmatrix}
        \frac{\partial^2}{\partial \pi^2_1} \bar{A}_i & \frac{\partial^2}{\partial \pi_1 \partial \pi_2} \bar{A}_i & \cdots & \frac{\partial^2}{\partial \pi_1 \partial \pi_N} \bar{A}_i\\
        \frac{\partial^2}{\partial \pi_2 \partial \pi_1} \bar{A}_i & \frac{\partial^2}{ \partial \pi^2_2} \bar{A}_i & \cdots & \frac{\partial^2}{\partial \pi_2 \partial \pi_N} \bar{A}_i\\
        \vdots & \vdots & \ddots & \vdots\\
        \frac{\partial^2}{\partial \pi_N \partial \pi_1} \bar{A}_i & \frac{\partial^2}{\partial \pi_N \partial \pi_2} \bar{A}_i & \cdots & \frac{\partial^2}{\partial \pi^2_N} \bar{A}_i
        \end{bmatrix}\\
        &= 2\bar{A}^3_i \begin{bmatrix}
        p^2_{1i} & p_{1i} p_{2i} & \cdots &  p_{1i} p_{Ni}\\
        p_{2i} p_{1i} & p^2_{2i} & \cdots &  p_{2i} p_{Ni}\\
        \vdots & \vdots & \ddots & \vdots\\
        p_{Ni} p_{1i} & p_{Ni} p_{2i} & \cdots &  p^2_{Ni}\\
        \end{bmatrix}\\
        &= 2\bar{A}^3_i \begin{bmatrix}p_{1i} \\ p_{2i}\\ \vdots \\ p_{Ni} \end{bmatrix} \begin{bmatrix}p_{1i} & p_{2i}& \cdots & p_{Ni} \end{bmatrix}\\
        &= 2\bar{A}^3_i \mathbf{P}_i \mathbf{P}^{T}_i.
    \end{aligned}
\end{align}
Here, $\mathbf{P}_i$ is the $i$-th column of the correlation matrix $\mathbf{P}$. Note that any matrix of the form $\mathbf{v}\mathbf{v}^{T}$ where $\mathbf{v} \in \mathbf{R}^{N}$, is positive semi-definite since:
\[ \mathbf{y}^{T} (\mathbf{v}\mathbf{v}^{T}) \mathbf{y} =  (\mathbf{v}^{T} \mathbf{y})^2 \geq 0, \forall \mathbf{y} \in \mathbf{R}^{N}.\]
Thus, the hessian $\mathbf{H}(\bar{A}_i)$ with respect to the scheduling probabilities $\bm{\pi}$ is positive semi-definite, which implies that $\bar{A}_i$ is a convex function of $\bm{\pi}$. Since the objective function in \eqref{eq:age_opt_prob_sr} is simply a weighted sum of the average AoIs and the sum of convex functions is convex, it is also a convex function of $\bm{\pi}$. Further, since the constraints of \eqref{eq:age_opt_prob_sr} are linear in $\bm{\pi}$, they are also trivially convex. Thus, the optimization problem \eqref{eq:age_opt_prob_sr} is convex as well. This proves the first part of \autoref{thm:sr_struc}.

Next, we apply KKT conditions to learn more about the structure of the optimal stationary randomized policy. To do so, we first formulate the Lagrangian function:
\[ \mathcal{L}(\bm{\pi}, \lambda, \bm{\mu}) = \sum_{i=1}^{N} \frac{w_i}{\sum_{j=1}^{N} \pi_j p_{ji}} + \lambda \bigg(\sum_{i=1}^{N}\pi_i - 1\bigg) - \sum_{i=1}^{N} \mu_i \pi_i.\]
Applying first-order KKT conditions, we see that:
\begin{equation}
    \frac{\partial}{\partial \pi_i} \mathcal{L}(\bm{\pi}^{*}, \lambda^{*}, \bm{\mu}^{*}) = - \sum_{j=1}^{N}  \frac{w_j p_{ij}}{(\sum_{k=1}^{N} \pi^{*}_k p_{kj})^2} + \lambda^{*} - \mu^{*}_i = 0, \forall i.
\end{equation}

Now, assume that an optimal scheduling policy $\bm{\pi}^{*}$ is one that only schedules sources from the set $S$, where $S \subseteq [N]$ and $[N]$ is the set of all sources. Then, $\pi^{*}_i > 0, \forall i \in S$. Applying complementary slackness for this policy $\bm{\pi}^{*}$, we get that $\mu_i = 0, \forall i \in S$. Combining this with the first order condition above, we get:
\begin{equation}
    \sum_{j=1}^{N}  \frac{w_j p_{ij}}{(\sum_{k=1}^{N} \pi^{*}_k p_{kj})^2} = \lambda, \forall i \in S.
\end{equation}
However, from \autoref{thm:avg_AoI}, we know that the average AoI of source $i$ under a policy $\bm{\pi}$ is given by $\frac{1}{\sum_{k=1}^{N} \pi_k p_{ki}}$. Thus, we can simplify the above equation to:
\begin{equation}
\label{eq:sys1}
    \sum_{j=1}^{N} p_{ij} w_j \bar{A}^2_j  = \lambda, \forall i \in S.
\end{equation}
Here, $\bar{A}_i$ denotes the average AoI of source $i$ under the policy $\bm{\pi}^{*}.$ Since $\bm{\pi}^{*}$ is an optimal policy, the total utilization of the channel should be 100\%. Thus, 
\begin{equation}
\label{eq:sys2}
    \sum_{i \in S} \pi^{*}_i = 1.
\end{equation} 
Observe that equations \eqref{eq:sys1} and \eqref{eq:sys2} together form a system of $|S|+1$ equations in $|S|+1$ variables. Solving this system gives us the values of $\pi^{*}_i, \forall i  \in S$. This completes our proof.

\section{Proof of Theorem~\ref{thm:sr_2opt}}
\label{pf:sr_2opt}
Consider a general scheduling policy $\bm{\pi}$, that is not necessarily stationary randomized. We will restrict ourselves to policies that achieve finite average AoI for each source $i$, since we are interested in minimizing AoI and can safely ignore oplicies for which the average AoI is unbounded. Thus, we will assume that all expressions involving limits are well-defined throughout the proof.

Let $I_{k,i}$ denote the $k$-th inter-arrival time between update deliveries regarding source $i$ to the base station. First, the average AoI of source $i$ under this policy is given by:
\begin{align}
    \begin{aligned}
        \bar{A}_i &\triangleq \lim_{T \rightarrow \infty} \bigg[ \frac{\sum_{t=1}^{T} A_i(t)}{T}  \bigg]
    = \lim_{K \rightarrow \infty}  \bigg[ \frac{\sum_{k=1}^{K}  I^2_{k,i} + I_{k,i}}{2(\sum_{k=1}^{K} I_{k,i})} \bigg].
    \end{aligned}
\end{align}
Next, define the following three three empirical average quantities:
\begin{align}
\begin{aligned}
    \hat{I}_i &\triangleq \frac{\sum_{k=1}^{K} I_{k,i}}{K},\\
    \hat{I}^{(2)}_i &\triangleq \frac{\sum_{k=1}^{K} I^{2}_{k,i}}{K},\\
    \hat{Var}_i &\triangleq \hat{I}^{(2)}_i - (\hat{I}_i)^2.
\end{aligned}
\end{align}
Then, using these definitions, we can simplify the expression for $\bar{A}_i$ as follows:
\begin{equation}
    \bar{A}_i = \lim_{K \rightarrow \infty}  \bigg[ \frac{1}{2} + \frac{\hat{I}^{(2)}_i}{2 \hat{I}_i}  \bigg] = \lim_{K \rightarrow \infty}  \bigg[ \frac{1}{2} + \frac{(\hat{I}_i)^2 + \hat{Var}_i}{2 \hat{I}_i}  \bigg].
\end{equation}
Using the Cauchy-Schwarz inequality, it is easy to see that $\hat{Var}_i \geq 0$. Thus, we can lower-bound the average AoI for source $i$ by:
\begin{equation}
\label{eq:lb1}
    \bar{A}_i \geq \frac{1}{2} + \lim_{K \rightarrow \infty} \frac{\hat{I}_i}{2}  .
\end{equation}

Now, define $f_j$ to be the fraction of time that the policy $\bm{\pi}$ schedules a source $j$ on average. Further, define $r_j$ to be the fraction of time that the base-station received a delivery about source $j$. Since there are correlated updates, $r_j \geq f_j$. Let $u_j(t)$ be an indicator variable that denotes whether policy $\bm{\pi}$ chooses source $j$ in time-slot $t$ or not and $X_{ji}(t)$ denote whether source $j$ has a correlated update about source $i$ at time $t$. Then,  
\begin{align}
\begin{aligned}
    r_i &\triangleq \lim_{T \rightarrow \infty} \frac{\sum_{t=1}^{T} \sum_{j=1}^{N} u_j(t) X_{ji}(t) }{T}.
\end{aligned}
\end{align}
Let the set of time-slots in which source $j$ is scheduled be $T_j = \{t: u_j(t) = 1 \}$. Using this definition, we rewrite the equation above as:
\begin{align}
\begin{aligned}
    r_i &= \lim_{T \rightarrow \infty}  \sum_{j=1}^{N} \frac{\sum_{ t \in T_j } X_{ji}(t)}{T} \\
    &= \lim_{T \rightarrow \infty}  \sum_{j=1}^{N} \frac{|T_j|}{T} \frac{\sum_{ t \in T_j } X_{ji}(t)}{|T_j|}.
\end{aligned}
\end{align}
Note that $\lim\limits_{T \rightarrow \infty} \frac{|T_j|}{T}$ is simply $f_j$. Further, if $f_j > 0$, then as $T \rightarrow \infty$, $|T_j|$ must also go to infinity and we can apply the law of large numbers to get $\lim\limits_{T \rightarrow \infty} \frac{\sum_{ t \in T_j } X_{ji}(t)}{|T_j|} = p_{ji}$. If $f_j = 0$, the law of large numbers cannot be applied. However, it is easy to see that $\lim\limits_{T \rightarrow \infty} \frac{\sum_{ t \in T_j } X_{ji}(t)}{T} = 0 = f_j p_{ji}$. Thus, the expression for $r_i$ simplifies to:
\begin{equation}
\label{eq:r1}
    r_i = \sum_{j=1}^{N} f_j p_{ji}.
\end{equation}

Another way to calculate $r_i$ is to consider the following limit:
\begin{equation}
    r_i = \lim_{K \rightarrow \infty}\frac{K}{\sum_{k=1}^{K} I_{k,i}}.
\end{equation}
Thus, if $\lim\limits_{K \rightarrow \infty} \hat{I}_i$ is well defined, then $r_i$ is also given by:
\begin{equation}
\label{eq:r2}
    r_i = \frac{1}{\lim\limits_{K \rightarrow \infty} \hat{I}_i}.
\end{equation}
Combining \eqref{eq:r1} and \eqref{eq:r2}, we get:
\begin{equation}
\label{eq:lb2}
    \lim\limits_{K \rightarrow \infty} \hat{I}_i = \frac{1}{\sum_{j=1}^{N} f_j p_{ji}}.
\end{equation}
Combining \eqref{eq:lb1} and \eqref{eq:lb2}, we get:
\begin{equation}
    \label{eq:lb3}
    \sum_{i=1}^{N} w_i \bar{A}_i \geq  \sum_{i=1}^{N} \frac{w_i}{2} + \frac{1}{2} \sum_{i=1}^{N} \frac{w_i}{\sum_{j=1}^{N} f_j p_{ji}}.
\end{equation}
Note that the frequencies of transmission for each source $f_j$ satisfy the following two constraints: $\sum_{j=1}^{N} f_j \leq 1$ and $f_j \geq 0, \forall j$. Thus, we can further lower-bound the weighted-sum average AoI by minimizing the RHS in \eqref{eq:lb3} over all choices of transmission frequencies:
\begin{equation}
    \label{eq:lb4}
    \sum_{i=1}^{N} w_i \bar{A}_i \geq  \sum_{i=1}^{N} \frac{w_i}{2} + \frac{1}{2} \sum_{i=1}^{N} \frac{w_i}{\sum_{j=1}^{N} f^{*}_j p_{ji}}.
\end{equation}
Here $\bm{f}^{*}$ is the solution to the following optimization problem:
\begin{align}
\begin{aligned}
\underset{\bm{\bm{f}}}{\operatorname{argmin}} & \sum_{i=1}^{N} \frac{w_i}{\sum_{j=1}^{N} f_j p_{ji}} \\
\text{s.t. }& \sum_{i=1}^{N} f_i \leq 1,\\
& f_i \geq 0, \forall i \in [N].
\end{aligned}
\end{align}
Observe that this optimization problem is identical to the \eqref{eq:age_opt_prob_sr} which solves for the optimal stationary randomized policy $\bm{\pi}^{*}$. Thus, $\bm{f}^{*} = \bm{\pi}^{*}$ and we get: \begin{equation}
    \label{eq:lb5}
    \sum_{i=1}^{N} w_i \bar{A}_i \geq  \sum_{i=1}^{N} \frac{w_i}{2} + \frac{1}{2} \sum_{i=1}^{N} \frac{w_i}{\sum_{j=1}^{N} \pi^{*}_j p_{ji}}.
\end{equation}

Note that \eqref{eq:lb5} is true for any scheduling policy $\bm{\pi}$ with finite average AoI for each source. So, it is also true for the policy $\bm{\pi}^{opt}$ that achieves minimum weighted-sum average AoI. This is because a simple round-robin policy achieves finite AoI for each source, so the optimal policy which performs at least as well as the round-robin policy, must also have well-defined and bounded average AoI for each source. Let's denote the average AoI for source $i$ under the overall optimal policy $\bm{\pi}^{opt}$ by $\bar{A}^{opt}_i$ and the average AoI under the optimal stationary randomized policy $\bm{\pi}^{*}$ by $\bar{A}^{*}_i$. Then, we get:
\begin{equation}
    \label{eq:lb6}
    \sum_{i=1}^{N} w_i \bar{A}^{opt}_i \geq  \sum_{i=1}^{N} \frac{w_i}{2} + \frac{1}{2} \sum_{i=1}^{N} w_i \bar{A}^{*}_i.
\end{equation}
Dividing, both sides in \eqref{eq:lb6} by the LHS and multiplying by 2, we get:
\begin{align}
\begin{aligned}
2 - \frac{\sum_{i=1}^{N} w_i}{ \sum_{i=1}^{N} w_i \bar{A}^{opt}_i } &\geq \frac{ \sum_{i=1}^{N} w_i \bar{A}^{*}_i }{ \sum_{i=1}^{N} w_i \bar{A}^{opt}_i},\\
\implies \frac{ \sum_{i=1}^{N} w_i \bar{A}^{*}_i }{ \sum_{i=1}^{N} w_i \bar{A}^{opt}_i} &\leq 2
\end{aligned}
\end{align}
This completes our proof.

\section{Proof of Theorem~\ref{thm:qmw}}
\label{pf:qmw}
Consider the one-slot Lyapunov drift for the quadratic Lyapunov function described by \eqref{eq:lyap2}:
\begin{align}
\begin{aligned}
\mathbb{E}\bigg[ &L(t+1) - L(t) \bigg| \mathbf{A}(t) \bigg] = \mathbb{E}\bigg[ \sum_{i=1}^{N}  w_i \big(A^2_i(t+1) - A^2_i(t+1) \big) \bigg]\\
&= \sum_{i=1}^{N} w_i \bigg( 2A_i(t) + 1 - \sum_{j=1}^{N} \mathbb{E}\bigg[ u_j(t)X_{ji}(t) \big| A_i(t) \bigg] A_i(t) (A_i(t) + 2)   \bigg)\\
&= \sum_{i=1}^{N} w_i \bigg( 2A_i(t) + 1 - \sum_{j=1}^{N} p_{ji} \mathbb{E}\bigg[ u_j(t) \big| A_i(t) \bigg] A_i(t) (A_i(t) + 2)   \bigg).
\end{aligned}
\end{align}
Since the quadratic max-weight policy minimizes the Lyapunov drift in every time-slot, we can upper-bound its one-slot drift by that of any other scheduling policy. In particular, we choose the optimal stationary randomized scheduling policy $\bm{\pi}^{*}$, for which we know that $\mathbb{E}[u_j(t) | A_i(t)] = \pi^{*}_j, \forall i,j,t$. Using this we upper-bound the one-slot Lyapunov drift as follows:
\begin{align}
\label{eq:4opt1}
    \begin{aligned}
    \mathbb{E}&\bigg[ L(t+1) - L(t) \bigg| \mathbf{A}(t) \bigg]  \\
    &\leq \sum_{i=1}^{N} w_i \bigg( 2A_i(t) + 1 - \sum_{j=1}^{N} p_{ji} \pi^{*}_j A_i(t) (A_i(t) + 2)   \bigg) \\
    &= \sum_{i=1}^{N} w_i \bigg( 2A_i(t) + 1 - r_i A_i(t) (A_i(t) + 2)   \bigg)\\
    &= -\sum_{i=1}^{N} w_i r_i \bigg( A_i(t) - \frac{1}{r_i} + 1 \bigg)^2 + \sum_{i=1}^{N} w_i  \Bigg(r_i\bigg(\frac{1}{r_1} - 1\bigg)^2 + 1 \Bigg)
    \end{aligned}
\end{align}
Here, $r_i\triangleq \sum_{j=1}^{N} p_{ji} \pi^{*}_j$, ad before. Next, using the Cauchy Schwarz inequality we get:
\begin{equation}
    \Bigg(\sum_{i=1}^{N} w_i r_i \bigg( A_i(t) - \frac{1}{r_i} + 1 \bigg)^2\Bigg) \bigg( \sum_{i=1}^{N} \frac{w_i}{r_i}  \bigg) \geq
    \Bigg(  \sum_{i=1}^{N} w_i \bigg|  A_i(t) - \frac{1}{r_i} + 1  \bigg| \Bigg)^2. 
\end{equation}
Using the above inequality in 
\eqref{eq:4opt1}, we get:
\begin{align}
    \begin{aligned}
    \mathbb{E}&\bigg[ L(t+1) - L(t) \bigg| \mathbf{A}(t) \bigg]  \\
    &\leq -\Bigg(  \sum_{i=1}^{N} w_i \bigg|  A_i(t) - \frac{1}{r_i} + 1  \bigg| \Bigg)^2 \bigg( \sum_{i=1}^{N} \frac{w_i}{r_i}  \bigg)^{-1} \\ &+ \sum_{i=1}^{N} w_i  \Bigg(r_i\bigg(\frac{1}{r_i} - 1\bigg)^2 + 1 \Bigg).
    \end{aligned}
\end{align}
Define $\Delta\big(\mathbf{A}(t)\big) \triangleq \mathbb{E}[ L(t+1) - L(t) | \mathbf{A}(t) ]$. Rearranging, we get:
\begin{align}
    \begin{aligned}
    \Bigg(  \sum_{i=1}^{N}& w_i \bigg|  A_i(t) - \frac{1}{r_i} + 1  \bigg| \Bigg)^2 \leq  -\bigg( \sum_{i=1}^{N} \frac{w_i}{r_i}  \bigg) \Delta\big(\mathbf{A}(t)\big) \\ &+  \bigg( \sum_{i=1}^{N} \frac{w_i}{r_i} \bigg) \sum_{i=1}^{N} w_i  \Bigg(r_i\bigg(\frac{1}{r_i} - 1\bigg)^2 + 1 \Bigg).
    \end{aligned}
\end{align}
Taking the expectation and summing over time, we get:
\begin{align}
\begin{aligned}
    \sum_{t=1}^{T} \mathbb{E}\Bigg[ \bigg(  \sum_{i=1}^{N}& w_i \bigg|  A_i(t) - \frac{1}{r_i} + 1  \bigg| \bigg)^2 \Bigg] \leq  -\bigg( \sum_{i=1}^{N} \frac{w_i}{r_i}  \bigg)\sum_{t=1}^{T} \mathbb{E}\bigg[ \Delta\big(\mathbf{A}(t)\big) \bigg]\\
    &+ T\bigg( \sum_{i=1}^{N} \frac{w_i}{r_i} \bigg) \sum_{i=1}^{N} w_i  \Bigg(r_i\bigg(\frac{1}{r_i} - 1\bigg)^2 + 1 \Bigg).
\end{aligned}
\end{align}
Dividing by $T$ and applying Jensen's inequality, we get:
\begin{align}
\begin{aligned}
    \frac{1}{T}&\sum_{t=1}^{T} \mathbb{E}\Bigg[ \bigg(  \sum_{i=1}^{N} w_i \bigg|  A_i(t) - \frac{1}{r_i} + 1  \bigg| \bigg) \Bigg]^2 \leq \\& -\bigg( \sum_{i=1}^{N} \frac{w_i}{r_i}  \bigg)\sum_{t=1}^{T} \mathbb{E}\bigg[ \frac{L(T+1) - L(1)}{T} \bigg]
    \\&+ \bigg( \sum_{i=1}^{N} \frac{w_i}{r_i} \bigg) \sum_{i=1}^{N} w_i  \Bigg(r_i\bigg(\frac{1}{r_i} - 1\bigg)^2 + 1 \Bigg)\\
    &\leq \bigg( \sum_{i=1}^{N} \frac{w_i}{r_i}  \bigg)\sum_{t=1}^{T} \mathbb{E}\bigg[ \frac{L(1)}{T} \bigg] + \bigg( \sum_{i=1}^{N} \frac{w_i}{r_i} \bigg) \sum_{i=1}^{N} w_i  \Bigg(r_i\bigg(\frac{1}{r_i} - 1\bigg)^2 + 1 \Bigg).
\end{aligned}
\end{align}
Applying Jensen's inequality again, and using the fact that $r_i \leq 1 \forall i$, we get:
\begin{align}
\begin{aligned}
    \mathbb{E}&\Bigg[ \frac{1}{T}\sum_{t=1}^{T}  \bigg(  \sum_{i=1}^{N} w_i \bigg|  A_i(t) - \frac{1}{r_i} + 1  \bigg| \bigg) \Bigg]^2  \\
    &\leq \bigg( \sum_{i=1}^{N} \frac{w_i}{r_i}  \bigg)\sum_{t=1}^{T} \mathbb{E}\bigg[ \frac{L(1)}{T} \bigg] + \bigg( \sum_{i=1}^{N} \frac{w_i}{r_i} \bigg) \sum_{i=1}^{N} w_i  \Bigg(r_i\bigg(\frac{1}{r_i} - 1\bigg)^2 + 1 \Bigg)\\
    &\leq \bigg( \sum_{i=1}^{N} \frac{w_i}{r_i}  \bigg)\sum_{t=1}^{T} \mathbb{E}\bigg[ \frac{L(1)}{T} \bigg] + \bigg( \sum_{i=1}^{N} \frac{w_i}{r_i} \bigg) \sum_{i=1}^{N} w_i  \bigg(r_i + \frac{1}{r_i} - 1\bigg)\\
    &\leq \bigg( \sum_{i=1}^{N} \frac{w_i}{r_i}  \bigg)\sum_{t=1}^{T} \mathbb{E}\bigg[ \frac{L(1)}{T} \bigg] + \bigg( \sum_{i=1}^{N} \frac{w_i}{r_i} \bigg) \sum_{i=1}^{N} \frac{w_i}{r_i}.
\end{aligned}
\end{align}
Taking the square-root of the inequality above and using the fact $L(1)$ is a constant, we take the limit as $T$ goes to infinity to get:
\begin{align}
    \begin{aligned}
        \lim_{T \rightarrow \infty} \mathbb{E}&\Bigg[ \frac{1}{T}\sum_{t=1}^{T}  \bigg(  \sum_{i=1}^{N} w_i \bigg|  A_i(t) - \frac{1}{r_i} + 1  \bigg| \bigg) \Bigg] \leq \bigg(\sum_{i=1}^{N} \frac{w_i}{r_i}\bigg).
    \end{aligned}
\end{align}
This inequality can be further simplified to:
\begin{equation}
    \lim_{T \rightarrow \infty} \mathbb{E}\Bigg[  \frac{1}{T}\sum_{t=1}^{T} \sum_{i=1}^{N} w_i A_i(t) \Bigg] \leq 2\bigg(\sum_{i=1}^{N} \frac{w_i}{r_i}\bigg).
\end{equation}
However, note that $r_i = \sum_{j=1}^{N} p_{ji} \pi^{*}_j$. Thus, we get:
\begin{equation}
    \sum_{i=1}^{N} w_i \bar{A}^{qmw}_i \leq 2\sum_{i=1}^{N} w_i \bar{A}^{*}_i,
\end{equation}
where $\bar{A}^{*}_i$ is the average AoI of source $i$ under the optimal stationary randomized policy $\bm{\pi}^{*}$. Since we have already shown factor-2 optimality of $\bm{\pi}^{*}$ in Appendix~\ref{pf:sr_2opt}, this completes the proof.

\section{Proof of Theorem~\ref{thm:mw_4opt}}
\label{pf:mw_4opt}
We first rewrite the correlated AoI evolution \eqref{eq:AoIEvolution} below:
\begin{equation}
    A_i(t+1) = A_i(t) + 1 - A_i(t) \sum_{j=1}^{N} u_j(t) X_{ji}(t), \forall i,t.
\end{equation}
Here, $u_j(t)$ indicates whether source $j$ transmits in time-slot $t$ or not and $X_{ji}(t)$ denotes whether source $j$ receives information about source $i$ at time-stop $t$ or not. Using this evolution and the definition of the Lyapunov function \eqref{eq:lyapunov}, we calculate the one-slot Lyapunov drift:
\begin{align}
\begin{aligned}
\mathbb{E}\bigg[ L(t+1) &- L(t) \bigg| \mathbf{A}(t) \bigg] = \mathbb{E}\bigg[ \sum_{i=1}^{N} \alpha_i - \\ &\sum_{i=1}^{N} \alpha_i A_i(t) \sum_{j=1}^{N} u_j(t)X_{ji}(t) \bigg| \mathbf{A}(t) \bigg]\\
&= \sum_{i=1}^{N} \alpha_i - \sum_{i=1}^{N}\alpha_i A_i(t) \sum_{j=1}^{N} \mathbb{E}\big[u_j(t) | \mathbf{A}(t)\big] p_{ji}.
\end{aligned}
\end{align}
Since the max-weight policy attempts to minimize the one-slot Lyapunov drift in every time-slot, the corresponding drift of any other policy must be greater than or equal to that of the max-weight policy. So, we can upper-bound the drift of max-weight by the drift of the optimal stationary randomized policy. Note that for the optimal randomized policy $\bm{\pi}^{*}$, the following holds:  \[\mathbb{E}\big[u_j(t)| \mathbf{A}(t)\big] = \pi^{*}_j, \forall j,t \]
since scheduling decisions are independent of the AoI values and across time-slots. Using this, we now upper-bound the one-slot drift of the max-weight policy as follows:
\begin{equation}
\label{eq:drift1}
    \mathbb{E}\bigg[ L(t+1) - L(t) \bigg| \mathbf{A}(t) \bigg] \leq \sum_{i=1}^{N} \alpha_i - \sum_{i=1}^{N} \sum_{j=1}^{N} \alpha_i \pi^{*}_j p_{ji} A_i(t).
\end{equation}
Taking expectations and summing \eqref{eq:drift1} from $t=1$ to $T$, we get:
\begin{equation}
\label{eq:drift2}
    \mathbb{E} \bigg[L(T+1) - L(1) \bigg] \leq T \sum_{i=1}^{N} \alpha_i - \sum_{i=1}^{N} \Bigg( \mathbb{E}\bigg[ \sum_{t=1}^{T} A_i(t) \bigg] \sum_{j=1}^{N} \alpha_i \pi^{*}_j p_{ji} \Bigg).
\end{equation}
Dividing \eqref{eq:drift2} by $T$, and re-arranging the terms, we get:
\begin{equation}
   \frac{1}{T} \sum_{i=1}^{N} \Bigg( \mathbb{E}\bigg[ \sum_{t=1}^{T} A_i(t) \bigg] \sum_{j=1}^{N} \alpha_i \pi^{*}_j p_{ji} \Bigg) \leq \sum_{i=1}^{N} \alpha_i + \frac{\mathbb{E}[L(1) - L(T+1)]}{T}. 
\end{equation}
Since $L(T+1) \geq 0$ and $\alpha_i = \frac{w_i}{\sum_{j=1}^{N} \pi^{*}_j p_{ji}}$, we get:
\begin{equation}
   \frac{1}{T} \sum_{i=1}^{N}  \mathbb{E}\bigg[ \sum_{t=1}^{T} w_i A_i(t) \bigg] \leq \sum_{i=1}^{N} \frac{w_i}{\sum_{j=1}^{N} \pi^{*}_j p_{ji}} + \frac{\mathbb{E}[L(1)]}{T}. 
\end{equation}
Taking the limit as $T$ goes to infinity, we get the following:
\begin{equation}
\label{eq:driftf}
    \sum_{i=1}^{N} w_i \bar{A}^{mw}_i \leq \sum_{i=1}^{N} w_i \bar{A}^{*}_i.
\end{equation}
Here $\bar{A}^{mw}_i$ is the average AoI of source $i$ under the max-weight policy while $\bar{A}^{*}_i$ is the average AoI of source $i$ under the optimal stationary randomized policy $\bm{\pi}^{*}$.

Note that we had already proved the factor-2 optimality of $\bm{\pi}^{*}$. Thus, \eqref{eq:driftf} is sufficient to obtain the same factor-2 optimality for the max-weight policy as well. This completes the proof.

\section{Proof of Theorem~\ref{thm:scaling}}
\label{pf:scaling}
Consider $N$ sources with the correlation matrix $\mathbf{P}$. Given a correlation threshold $p>0$, construct a directed graph $\mathcal{G}$ that represents pairs of source with correlation higher than the threshold.

Further, assume that the set $S \subseteq [N]$ is a minimum size vertex covering of the graph $\mathcal{G}$. Recall that we defined the notion of vertex covering for directed graphs in Sec.~\ref{sec:scaling}. Let the size of this minimum vertex covering be denoted by $N_{cov}$.

We will show that under a specific scheduling policy $\pi$, the average AoI of every source is upper-bounded by $\frac{N_{cov}}{p}$. Consequently, the weighted sum of average AoIs under an optimal policy will also be upper-bounded by $\frac{N_{cov}}{p}$, since we assumed equal weights. The scheduling policy $\pi$ we analyze is round-robin scheduling of sources in the covering set $S$. This is a cyclic policy of length $N_{cov}$ time-slots where each source in the covering set $S$ gets scheduled once, after which the scheduling pattern repeats every $N_{cov}$ time-slots.

Before we analyze the performance of this policy, we present a lemma that discusses the monotonicity of AoI with the correlation parameters.
\begin{lemma}
\label{lem:monotone}
Consider $N$ sources with two different correlation matrices $\mathbf{P}$ and $\mathbf{P}'$. If $p_{ij} \geq p'_{ij}, \forall i,j \in [N]$ then under a fixed scheduling policy $\pi$, the following holds:
\begin{equation}
    \bar{A}_i \leq \bar{A}'_i, \forall i \in [N].
\end{equation}
Here, $\bar{A}_i$ is the average AoI of source $i$ under policy $\pi$ with the correlation matrix $\mathbf{P}$, while $\bar{A}'_i$ is the average AoI of source $i$ under policy $\pi$ with the correlation matrix $\mathbf{P}'$. 
\end{lemma}
\begin{proof}
The proof is easy to see via a stochastic dominance argument. Let $X_{ji}(t)$ be an indicator variable denoting whether $j$ had information about $i$ at time-slot $t$ given the correlation matrix $\bm{P}$, and likewise $X'_{ji}(t)$ for the matrix $\bm{P}'$. Then, for all pairs $(i,j)$ and for all time-slots $t$, $X_{ij}(t) \sim Bern(p_{ij})$ and $X'_{ij}(t) \sim Bern(p'_{ij})$, where $p_{ij} \geq p'_{ij}$. Thus, $X_{ij}(t) \geq_{st.} X'_{ij}(t)$.

Now, the AoI evolution \eqref{eq:AoIEvolution} can be rewritten as:
\begin{align}
\label{eq:coup}
\begin{aligned}
    A_i(t+1) &= A_i(t) + 1 - A_i(t) \sum_{j=1}^{N} u_j(t) X_{ji}(t), \text{ and }\\
    A'_i(t+1) &= A'_i(t) + 1 - A'_i(t) \sum_{j=1}^{N} u'_j(t) X'_{ji}(t), \forall i,t.
\end{aligned}
\end{align}
Since we have fixed the policy $\pi$, the scheduling decisions remain the same for both correlation matrices, i.e. $u_j(t) = u'_j(t), \forall t$. Setting $A_i(1) = A'_i(1)$, \eqref{eq:coup} implies that:
\begin{equation}
    A_i(t) \leq_{st.} A'_i(t).
\end{equation}
This, in turn, further implies that:
\begin{equation}
   \frac{1}{T} \sum_{t=1}^{T} A_i(t) \leq_{st.} \frac{1}{T} \sum_{t=1}^{T} A'_i(t), \forall i,t.
\end{equation}
Using the equation above and the definition of average AoI \eqref{eq:avg_AoI_def}, we can conclude that:
\begin{equation}
    \bar{A}_i \leq_{st.} \bar{A}'_i, \forall i \in [N].
\end{equation}
If the average AoI limits exist, then $\bar{A}_i$ and $\bar{A}'_i$ are just constants and the stochastic dominance becomes a simple inequality. This completes the proof.
\end{proof}

The lemma above shows a rather simple result: AoI improves with correlation. To upper-bound the average AoI of sources under policy $\pi$, it is sufficient to analyze AoI for a new correlation matrix $\mathbf{P}'$, which is defined as:
\begin{equation}
    p'_{ij} = \begin{cases}
    p, &\text{ if }p_{ij} \geq p.\\
    0, &\text{ otherwise.}
    \end{cases}
\end{equation}
Clearly, $\mathbf{P}'$ is element-wise smaller than $\mathbf{P}$, so we can apply Lemma~\ref{lem:monotone}. Now, to analyze AoI of a source $i$ under the policy $\pi$ with the correlation matrix $\mathbf{P}'$, we consider two cases. 

\textbf{Case 1:} Source $i$ is in the vertex covering set $S$. In this case, note that this source is guaranteed to be scheduled at least once in every $N_{cov}$ time-slots due to way round-robin scheduling works. It is possible that the base station might receive a correlated update about $i$ from some other source, even between two consecutive updates from $i$ that are $N_{cov}$ time-slots apart. However, since we are trying to upper-bound the AoI, we can safely ignore these updates. Thus, the average AoI of source $i$ can be upper-bounded by:
\begin{align}
\label{eq:c1}
    \begin{aligned}
    \bar{A}_i \leq \frac{\sum_{k=1}^{N_{cov}} k}{N_{cov}} &\leq \frac{N_{cov}+1}{2}\\
    & \leq \frac{N_{cov}}{p}.
    \end{aligned}
\end{align}
The last inequality follows since $p \leq 1$ and $N_{cov} \geq 1$.

\textbf{Case 2:} Source $i$ is not in the vertex covering set $S$. By the definition of a vertex covering, there must be at least one source $j \in S$ such that the edge $(j,i)$ is in the graph $\mathcal{G}$ or alternatively, that $p'_{ji} = p$. This means that whenever the scheduling policy schedules source $j$, the base station also receives an update regarding source $i$ with probability $p$. Note that the scheduling policy $\pi$ schedules source $j$ once every $N_{cov}$ time-slots. Then, the time intervals between two successful {correlated update deliveries} regarding source $i$ by source $j$ are $N_{cov} I_{1}, N_{cov} I_2, ..., N_{cov} I_K$ where $I_1, I_2, ..., I_K$ are i.i.d. geometrically distributed random variables with parameter $p$. It is possible that there are other sources also correlated with source $i$ that deliver correlated updates to the base-station between two consecutive correlated updates from source $j$. For the sake of upper-bounding the AoI we can safely ignore any such updates. Then, the average AoI of source $i$ can be upper-bounded by:
\begin{align}
    \begin{aligned}
    \bar{A}_i &\leq  \lim_{K \rightarrow \infty} \frac{ \sum_{k=1}^{K} (1 + 2 + ... + N_{cov}I_k)  }{ \sum_{k=1}^{K} N_{cov}I_k }\\
    &\leq\lim_{K \rightarrow \infty} \frac{\sum_{k=1}^{K} N^2_{cov}I^2_k + N_{cov}I_k }{2 \sum_{k=1}^{K} N_{cov}I_k}\\
    &\leq \frac{1}{2} + N_{cov} \lim_{K \rightarrow \infty}\frac{ \sum_{k=1}^{K} I^2_{k} }{2\sum_{k=1}^{K} I_k}.
    \end{aligned}
\end{align}
Applying the law of large numbers and using the moments of a geometric random variable, we get:
\begin{align}
\label{eq:c2}
\begin{aligned}
    \bar{A}_i \leq &\frac{1}{2} + N_{cov} \frac{(2-p)/p^2}{2/p}\\
    &\leq \frac{N_{cov}}{p} - \frac{N_{cov} - 1}{2}\\
    &\leq \frac{N_{cov}}{p}.
\end{aligned}
\end{align}
The last inequality follows since a vertex covering must have at least one vertex, i.e. $N_{cov} \geq 1$.

Together, \eqref{eq:c1} and \eqref{eq:c2} imply that the average AoI $\bar{A}_i$ for every source $i$ under the vertex cover round-robin policy $\pi$ is upper-bounded by $\frac{N_{cov}}{p}$. Clearly, the performance of the policy that achieves minimum weighted-sum average AoI must be better than that of $\pi$. This implies:
\begin{align}
    \begin{aligned}
        \sum_{i=1}^{N} w_i \bar{A}^{opt}_i &\leq \sum_{i=1}^{N} w_i \bar{A}_i \\
        & \leq \frac{N_{cov}}{p}.
    \end{aligned}
\end{align}
The last inequality follows since we have set all weights $w_i$ to be equal to $\frac{1}{N}$. This completes the proof.

\section{Proof of Theorem~\ref{lem:compare}}
\label{pf:rr_maf}
We want to prove a lower bound on the performance of the max-AoI-first policy $\bm{\pi}^{u}$. To do so, we start with $\bm{A}(1)$ such that $A_i(1) = i$. Clearly, the policy does not schedule source $1$ at time $t=1$ since it does not have the maximum AoI. In fact, since sources $2,...,N$ are uncorrelated, source $1$ only has a chance to get scheduled once its AoI reaches $N$. This is because max-AoI-first over the set of uncorrelated sources $\{2,...N\}$ is simply the round-robin policy and the round-robin policy over $N-1$ sources always has at least one source with AoI $N-1$ in any time-slot. Suppose at some time-slot $t$ the AoI of source $1$ hits $N$ and so it gets scheduled, breaking the round-robin phase. Then, in the next time-slot, the AoI of source $1$ will go down to $1$ while the AoI of all other sources will be at least $1$ and possibly greater (if $1$ failed to send a correlated update). Assuming a tie-breaking rule that always prefers sources with higher indices, max-AoI-first again starts scheduling sources $2,...,N$ since source $1$ has the smallest AoI. Again, source $1$ will only be scheduled once its AoI reaches $N$ and during this period $\bm{\pi}^{u}$ will simply be round-robin.

Since all the sources $2,...,N$ are correlated with source $1$ with probability $p$, the AoI of source $1$ during each round-robin phase grows as a geometric random variable with parameter $p$. Define $f_i$ to be the fraction of time that the policy $\bm{\pi}^{u}$ schedules a source $i$ on average. Then, we can upper-bound $f_1$ by looking at how often a geometric random variable with the parameter $p$ is smaller than $N$.
\begin{equation}
\label{eq:pf1}
    f_1 \leq 1 - \mathbb{P}(Geo(p) \leq N-1) = (1-p)^{N-1}. 
\end{equation}
Note that the policy $\bm{\pi}^{u}$ is equivalent to round-robin over $2,...,N$ the rest of the time, when $1$ is not being scheduled. Thus, $f_i = f, \forall i \neq 1$. Further, since the policy $\bm{\pi}^{u}$ never idles, we know that:
\begin{equation}
\label{eq:pf2}
    f_1 + (N-1)f  = 1
\end{equation}
Together, \eqref{eq:pf1} and \eqref{eq:pf2} imply the following:
\begin{equation}
\label{eq:pf3}
    f \geq \frac{1-(1-p)^{N-1}}{N-1} \triangleq f_{min}. 
\end{equation}

Next, define $r_i$ to be the fraction of time that the base-station received a delivery about source $i$. In Appendix~\ref{pf:sr_2opt}, we show the following for every policy with well-defined average AoI:
\begin{equation}
     r_i = \sum_{j=1}^{N} f_j p_{ji}, \forall i \in [N].
\end{equation}
Putting in $f_i = f, \forall i \neq 1$ and using the correlation matrix $\mathbf{P}$, we get:
\begin{equation}
\label{eq:thp}
    r_i = \begin{cases}
    1-f(N-1)(1-p), &\text{ if }i=1,\\
    f\big(1- (N-1)p\big) + p, &\text{ otherwise }.
    \end{cases}
\end{equation}
We know from analysis in Appendix~\ref{pf:sr_2opt} that for any scheduling policy $\bm{\pi}$ the following holds:
\begin{equation}
\label{eq:pflb}
    \sum_{i=1}^N w_i \bar{A}_i \geq \frac{1}{2} \sum_{i=1}^N \frac{w_i}{r_i}.
\end{equation}

Applying the inequality \eqref{eq:pflb} to the uncorrelated max-weight policy $\bm{\pi}^{u}$ and putting in values of $r_i$ from \eqref{eq:thp}, we get:
\begin{align}
    \begin{aligned}
    \label{eq:cex2}
        \sum_{i=1}^{N} w_i \bar{A}^{u}_i &\geq \frac{1}{2N} \frac{1}{1-f(N-1)(1-p)} + \frac{N-1}{2N} \frac{1}{f(1-(N-1)p) + p} \\
        &\geq \frac{N-1}{2N} \frac{1}{f_{min}(1-(N-1)p) + p}.
    \end{aligned}
\end{align}
The last inequality above follows by assuming $p \geq \frac{1}{N-1}$ and using the fact that $\frac{1}{r_i}$ become monotone \textit{increasing} functions of $f$ so we can lower-bound the RHS by choosing the smallest possible value of $f$. Putting in the expression for $f_{min}$ from \eqref{eq:pf3}, and simplifying we get:
\begin{equation}
\label{eq:cex3}
    \sum_{i=1}^{N} w_i \bar{A}^{u}_i \geq \frac{(N-1)^2}{2N - (N+1)(1-p)^{N-1}}.
\end{equation}
This lower bound combined with the upper bound for the correlated max-weight policy completes the proof.

\section{Proof of Theorem~\ref{thm:rgg}}
\label{pf:rgg}
\begin{figure}
	\centering
	\includegraphics[width=0.8\linewidth]{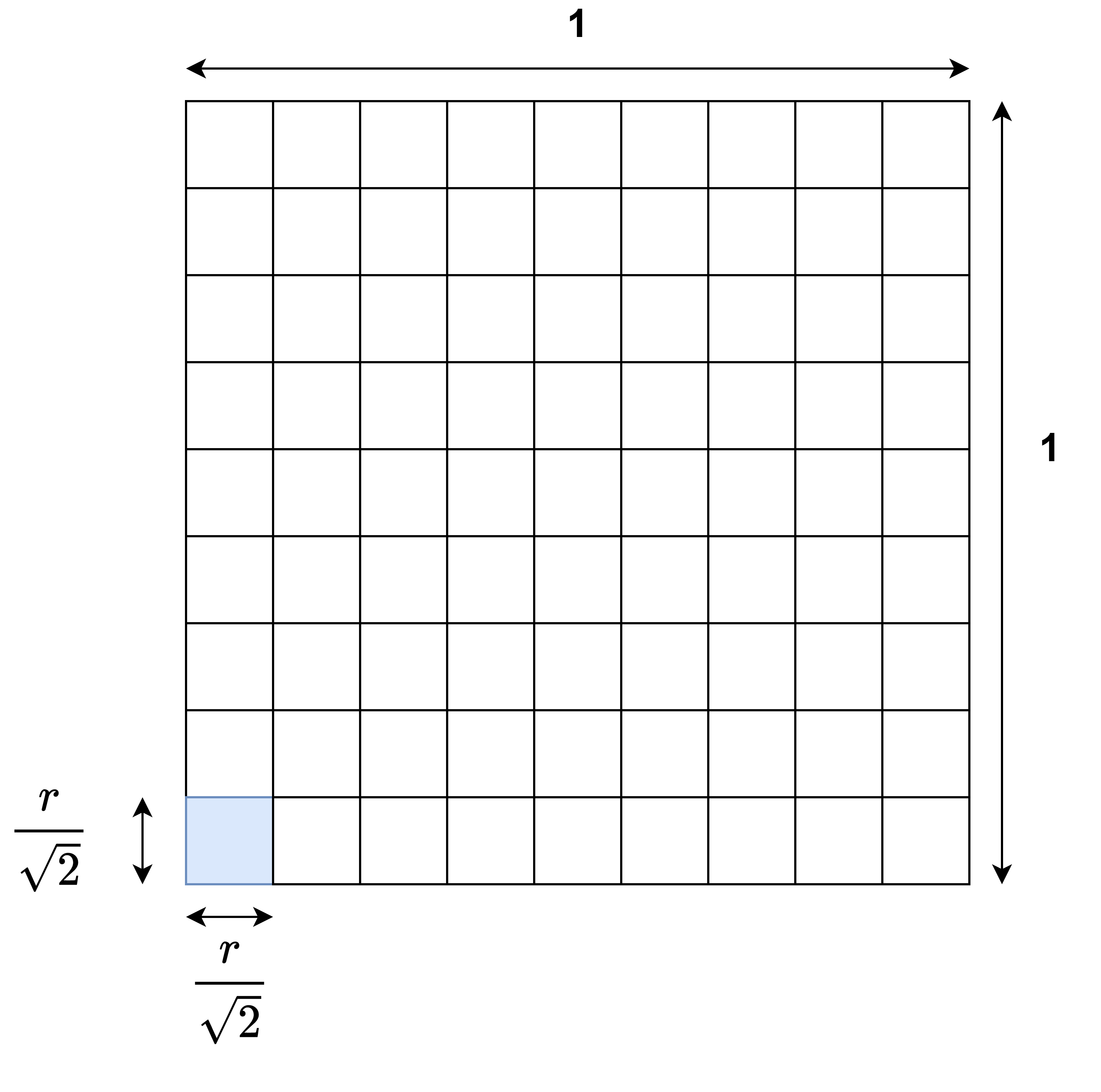}
	\caption{Grid of size $\frac{r}{\sqrt{2}}$ on the unit square.}
	\label{fig:grid}
	\vspace{-1mm}
\end{figure}
Consider a symmetric correlation matrix generated by creating a random geometric graph $\mathcal{G}(N,r)$ on the two dimensional unit square and setting correlation probabilities for neighbors to be $p$. Assume equal weights $w_i = \frac{1}{N}, \forall i$.

We will apply the vertex covering result from \autoref{thm:scaling} to this geometric graph setting. Divide the unit square into square cells of size $r/\sqrt{2} \times r /\sqrt{2}$ (see Figure~\ref{fig:grid}). For each cell on this grid, choose one source within the cell to be a member of the vertex covering set $S$. If there are no sources, ignore the cell and if there are more than one sources, pick one at random. Note that every source within one such cell is at most a distance $r$ away from any other source within the same cell. Thus, all source pairs $(i,j)$ within the same cell are connected on $\mathcal{G}(N,r)$ and must have correlation probabilities $p_{ij} = p_{ji} = p$. 

It is easy to see that the set $S$, which consists of at most one unique source from each cell, is a vertex covering on $\mathcal{G}(N,r)$. The size of the set $S$ is upper-bounded by the total number of cells. Since the area of each cell is $\frac{r^2}{2}$, the total number of cells required to cover the unit square is $\frac{2}{r^2}$. Further, since all correlation probabilities are equal to $p$, we can use \autoref{thm:scaling} on the graph $\mathcal{G}(N,r)$ with the correlation threshold $p$. This gives us:
\begin{equation}
   \sum_{i=1}^{N} w_i \bar{A}^{opt}_i \leq \frac{N_{cov}}{p} \leq \frac{2}{p r^2}.
\end{equation}
The first part of the inequality is a direct application of \autoref{thm:scaling} while the second part holds since we have a found a vertex covering of size at-most $\frac{2}{r^2}$ which is an upper-bound on the size of the minimum vertex covering $N_{cov}$. This completes our proof.

\section{Proof of Theorem~\ref{thm:equi_sr}}
\label{pf:equi_sr}
Consider the AoIs evolution:
\begin{equation}
    A_{i}(t+1) = A_i(t) + 1 - \bigg( \sum_{j=1}^{N} u_j(t) X_{ji}(t) \bigg) A_i(t), \forall i, t.
\end{equation}
Taking conditional expectation with respect to $A_i(t)$ we get:
\begin{equation}
    \mathbb{E}[A_i(t+1)| A_i(t)] =  A_i(t) + 1 - \bigg( \sum_{j=1}^{N} \mathbb{E} [u_j(t) | A_i(t)] p_{ji} \bigg) A_i(t), \forall i, t.
\end{equation}
Now, for a stationary randomized policy $\bm{\pi}$ we know that \[\mathbb{E} [u_j(t) | A_i(t)] = \pi_j, \forall j,i,t.\]
Thus, we get:
\begin{equation}
    \mathbb{E}[A_i(t+1)| A_i(t)] =  A_i(t) + 1 - \bigg( \sum_{j=1}^{N} \pi_j p_{ji} \bigg) A_i(t), \forall i, t.
\end{equation}
Denoting $\sum_{j=1}^{N} \pi_j p_{ji}$ by $r_i$, we get:
\begin{equation}
    \mathbb{E}[A_i(t+1)| A_i(t)] =  (1-r_i)A_i(t) + 1, \forall t \in \mathbb{N}.
\end{equation}

Observe that if $r_i = 0$ then $A_i(t)$ simply increases linearly with time and the average AoI is unbounded. This completes one part of the proof. For the remaining part, assume $r_i > 0$. 

Taking expectation again, we get:
\begin{equation}
\mathbb{E}\big[\mathbb{E}[A_i(t+1)| A_i(t)]\big] =  (1-r_i)\mathbb{E}\big[ \mathbb{E}[ A_i(t) | A_i(t-1)]\big] + 1, \forall t \in \mathbb{N}.
\end{equation}
Solving the recursion in the equation above we get:
\begin{equation}
    \mathbb{E}[A_i(t+1)] = (1-r_i)^{t} A_i(1) + \sum_{\tau=0}^{t-1} (1-r_i)^{\tau}.   
\end{equation}
Summing up the above for $t=0,...,T-1$ we get:
\begin{equation}
    \mathbb{E}\bigg[\sum_{t=1}^{T} A_i(t)\bigg] = \frac{1 - (1-r_i)^T}{r_i} A_i(1) + \frac{T}{r_i} - \frac{1-(1-r_i)^T}{r^2_i}.
\end{equation}
Dividing the equation above by $T$ and taking the limit as $T$ goes to infinity:
\begin{equation}
\label{eq:in_exp}
    \lim_{T \rightarrow \infty} \mathbb{E}\bigg[ \frac{\sum_{t=1}^{T} A_i(t)}{T} \bigg] = \frac{1}{r_i}.
\end{equation}
If we consider a sequence of random variables $\frac{\sum_{t=1}^{T} A_i(t)}{T}$, then \eqref{eq:in_exp} shows that this sequence converges in expectation to $\frac{1}{r_i}$. However, convergence in expectation implies convergence in probability. Thus, we get:
\begin{equation}
    \lim_{T \rightarrow \infty}  \frac{\sum_{t=1}^{T} A_i(t)}{T}  \overset{p}{=} \frac{1}{r_i}.
\end{equation}
Observe that the quantity on the LHS is the average AoI, so we have shown that:
\begin{equation}
    \bar{A}_i = \frac{1}{\sum_{j=1}^{N} \pi_j p_{ji}}.
\end{equation}
This completes our proof.

\end{document}